\documentclass{aa}
\usepackage{graphicx}
\usepackage{txfonts}
\usepackage{epstopdf}
\begin{document}

\title{Searching for faint companions with VLTI/PIONIER}

\subtitle{II. 92 main sequence stars from the Exozodi survey\thanks{Based on observations made with ESO Telescopes at the La Silla Paranal Observatory under programme IDs 089.C-0365 and 090.C-0526.}}

\titlerunning{Searching for faint companions with VLTI/PIONIER. II.}

\author{L.~Marion\inst{1}, O.~Absil\inst{1}\fnmsep\thanks{F.R.S.-FNRS Research Associate}, S.~Ertel\inst{2,3}, J.-B.~Le Bouquin\inst{3}, J.-C.~Augereau\inst{3}, N.~Blind\inst{4}, D.~Defr\`ere\inst{5}, J.~Lebreton\inst{6}, J.~Milli\inst{2,3}}

\authorrunning{Marion et al.}

\institute{D\'epartement d'Astrophysique, G\'eophysique et Oc\'eanographie, Universit\'e de Li\`ege, 17 All\'ee du Six Ao\^ut, 4000 Li\`ege, Belgium
\and 
European Southern Observatory, Alonso de Cordova 3107, Vitacura, Casilla 19001, Santiago 19, Chile
\and
Univ.\ Grenoble Alpes / CNRS, Institut de Plan\'etologie et d'Astrophysique de Grenoble (IPAG), F-38000 Grenoble, France
\and
Max-Planck-Institut f\"ur extraterrestrische Physik, Gie{\ss}enbachstra{\ss}e, 85741 Garching, Germany
\and
Department of Astronomy, University of Arizona, 993 N. Cherry Ave, Tucson, AZ 85721, USA
\and
Infrared Processing and Analysis Center, California Institute of Technology, Pasadena, CA 91125, USA
}

   \date{Received xxx; accepted xxx}

% \abstract{}{}{}{}{} 
% 5 {} token are mandatory
 
  \abstract
  % context heading (optional)
  % {} leave it empty if necessary  
   {The \textsc{Exozodi} survey aims to determine the occurrence rate of bright exozodiacal discs around nearby main sequence stars using infrared interferometry. Although the \textsc{Exozodi} survey targets have been carefully selected to avoid the presence of binary stars, the results of this survey can still be biased by the presence of unidentified stellar companions.}
  % aims heading (mandatory)
   {Using the PIONIER data set collected within the \textsc{Exozodi} survey in 2012, we aim to search for the signature of point-like companions around the \textsc{Exozodi} target stars.}
  % methods heading (mandatory)
   {We make use of both the closure phases and squared visibilities collected by PIONIER to search for companions within the $\sim$100~mas interferometric field of view. The presence of a companion is assessed by computing the goodness of fit to the data for a series of binary models with various separations and contrasts.}
  % results heading (mandatory)
   {Five stellar companions are resolved for the first time around five A-type stars: HD~4150, HD~16555, HD~29388, HD~202730, and HD~224392 (although the companion to HD~16555 was independently resolved by speckle interferometry while we were carrying out the survey). In the most likely case of main sequence companions, their spectral types range from A5V to K4V. Three of these stars were already suspected to be binaries from Hipparcos astrometric measurements, although no information was available on the companions themselves so far. In addition to debiasing the statistics of the \textsc{Exozodi} survey, these results can also be used to revise the fraction of visual binaries among A-type stars, suggesting that an extra $\sim$13\% A-type stars are visual binaries in addition to the ones detected in previous direct imaging surveys.}
  % conclusions heading (optional), leave it empty if necessary 
   {We estimate that about half the population of nearby A-type stars could be resolved as visual binaries using a combination of state-of-the-art interferometry and single-aperture imaging, and we suggest that a significant fraction of these binaries remains undetected to date.}


   \maketitle
%
%________________________________________________________________

\section{Introduction}

Historically, binary stars have been detected and studied in two main ways: (i) using the orbital motion around the centre of mass that can be measured with spectroscopy (or possibly with astrometry), or (ii) using direct (visual) observations. These two classes of techniques, which can be declined in various flavours, have discovery spaces that may or may not overlap depending on the considered target. In particular, the longest periods reachable by spectroscopy  frequently correspond to angular separations too short to be reached with classical imaging observations. This can lead to significant incompleteness when taking a census of binary stars. High angular resolution observations, such as those provided by infrared interferometry, may help bridge the gap between spectroscopy and classical imaging techniques.

Here, we aim to search for unknown companions around a specific sample of main sequence stars that was built in the context of an unbiased interferometric search for hot debris discs (also known as exozodiacal discs). The main purpose of this survey, presented in a companion paper \citep{Ertel14}\footnote{See also \citet{Absil13} for a description of the first part of the \textsc{Exozodi} survey, carried out in the northern hemisphere with the FLUOR instrument on the CHARA array.}, is to obtain statistically significant information on the occurrence of bright exozodiacal discs around nearby main sequence stars. Binary stars were removed from this sample to prevent contamination of the interferometric observations with spurious circumstellar emission, which would make faint exozodiacal disc detection much more difficult. This selection criterion also avoids gravitational interactions with nearby, massive bodies, which may induce biases by enhancing or disrupting the dusty discs around the target stars. The selection of single stars was based on the literature, which, as explained above, may not be complete regarding companions -- especially faint and close ones -- around these stars.

To make sure that the stars observed within the so-called \textsc{Exozodi} survey, carried out in the southern hemisphere with the Precision Integrated Optics Near Infared Experiment  \citep[PIONIER,][]{LeBouquin11} visitor instrument at the VLTI, are indeed single stars, we use the full information delivered by PIONIER (including closure phases) to systematically search for companions around all the stars observed within the survey. Doing so, we will be able to determine whether the small near-infrared excess detected around several stars within the survey are effectively due to an extended, mostly symmetric source, or to a point-like companion. This is of utmost importance in order to derive unbiased statistics on the occurrence rate of exozodiacal discs.

After detailing the stellar sample, the observations and data reduction in Sect.~\ref{sec:obsandred}, we present in Sect.~\ref{sec:searchcomp} the search method and the detection criterion. The results in terms of newly detected companions are given in that same section. A discussion of each new binary system is given in Sect.~\ref{sec:properties}. We finally discuss the implications of this study in terms of revised binary fractions and the PIONIER sensitivity to faint companions in Sect.~\ref{sec:discussion}.
%__________________________________________________________________

\section{Observations and data reduction}
\label{sec:obsandred}

The choice of the \textsc{Exozodi} stellar sample was driven by the goal to assess possible correlations between the presence of cold and hot dust around main sequence stars \citep{Ertel14}. The targets were selected from a magnitude limited sample of stars known to harbour cold dust in their outer planetary system. An equivalent sample of stars without any detectable cold dust emission (to within the sensitivity limit of far-infrared space-based observatories) was then selected, resulting in an all-sky sample of more than 200 stars. Because of the selection process, the overall sample is biased towards dusty stars (which are less frequent than non-dusty stars in our Galaxy at the sensitivity level of current far-infrared space observatories), although each of the two sub-samples is designed to be free from bias. The stars are more or less equally spread between three spectral type categories: A-type, F-type and G-K-type. All stars have been thoroughly checked for the presence of known (sub-)stellar companions in the literature; any star with a companion more massive than the planet-brown dwarf boundary and at a separation smaller than $5\arcsec$ was discarded. A more detailed description of the sample can be found in \citet{Ertel14}.

Ninety-two southern stars from the \textsc{Exozodi} sample were observed with the PIONIER visitor instrument operated on the 1.8-m Auxiliary Telescopes (AT) of the VLTI during four observing campaigns of three nights each from April to December 2012. The observations were conducted in H band on the most compact AT configuration (A1--B2--C1--D0), using the \textit{SMALL} spectral dispersion that splits the light into three spectral channels on the detector. Three consecutive observing blocks (OBs) were obtained for most stars, and care was taken to observe as many stars from each of the dusty and non-dusty categories. The optical path length (OPD) scans ranged from $60~\mu$m to $180~\mu$m depending on the baseline, and 1024 samples were read per scan to ensure a proper sampling of the fringes. Only the A and C outputs of the ABCD fringe coding scheme were read to speed up the readout.

We used the same data reduction scheme as presented in \citet{Ertel14}, based on the \texttt{pndrs} pipeline and including the specific calibration strategy developed for the \textsc{Exozodi} survey to optimise the accuracy of the squared visibilities. As a sanity check, we also carried out our study based on a more standard calibration scheme, where the transfer function is averaged over each individual CAL-SCI-CAL-SCI-CAL-SCI-CAL sequence with a weight inversely related to the time separation, as described in \citet{LeBouquin11}. The final results did not show any significant change.

%__________________________________________________________________

\section{Searching for companions}
\label{sec:searchcomp}

	\subsection{Principle of the search} \label{sub:principle}

To detect the presence of a companion, we use the closure phases and the squared visibilities in a combined way. Before explaining the details of the method used here, we will summarise the principle of the method used by \citet{Absil11}, which was also used as a first step in our analysis. The search for companions in \citet{Absil11} was based only on the closure phases. Indeed, the closure phase (CP) is known to be free of instrumental and atmospheric errors to the first order \citep{LeBouquin12}, and to be sensitive to asymmetries. For a centro-symmetric source, the CP will be strictly zero (or possibly a constant offset from zero due to imperfect CP calibration), while the presence of an off-axis companion will show up as a non-zero closure phase, which varies as a function of time and wavelength. \citet{Absil11} define a field of view and compute the CP associated with a binary model considering the primary star at the centre of the search region with an off-axis companion of various contrast $r$ starting at $r=0$, at each point $(x,y)$ of the field of view in the search region. The modelled CP is then compared to the observations to derive the $\chi^2$ goodness of fit for each binary model,
\begin{equation}
\chi^2_{\rm CP} = \sum\frac{({\rm CP}_{mod} - {\rm CP}_{data})^2}{\sigma^2_{\rm CP}} \; ,
\end{equation}
thereby creating a $\chi^2$ cube. The cube is then normalised (by division) so that its minimum equals 1, and collapsed along the contrast axis to keep only the best-fitting companion contrast (i.e.\ minimum $\chi^2$ value) at each position in the search region. The resulting $\chi^2$ map can then be used to derive the probability for the single-star model to adequately represent the data, based on the $\chi^2$ distribution under Gaussian noise assumption \citep[see][for details]{Absil11}. If this probability is below a pre-defined threshold, the single-star model can be rejected and the best-fit binary solution is then considered as statistically significant. In \citet{Absil11}, the threshold is fixed at a $3\sigma$ level, i.e.\ at a probability of $0.27\%$ under the Gaussian noise assumption.

In the present study, we noticed that the use of the sole CP leads to a large number of barely significant detections, with significance levels (i.e.\ signal-to-noise ratios) between 3 and 4$\sigma$. It was not obvious to discriminate whether these were true detections or false positives. As a consequence, in addition to the CP, we use the squared visibilities ($V^2$), which are expected to show deviations from the single-star model if a companion is present. As in \citet{Absil11}, we compute a binary model considering the primary star at the centre of the search region with an off-axis companion of various contrast $r$ (starting at $r<0$ as explained in Sect.~\ref{sub:detecriteria}) at each point $(x,y)$ of the search region. In the present case, we can safely assume that both the primary and the secondary stars are unresolved, since most of our targets have an angular diameter smaller than 1~mas \citep{Ertel14} and as the observations are carried out on the most compact AT configuration, which provides an angular resolution of the order of 5~mas. Then, we compute the CP and the $V^2$ for each model and derive a combined goodness of fit,
\begin{equation}
\chi^2 = \underbrace{\sum\frac{(V^2_{mod} - V^2_{data})^2}{\sigma^2_{V^2}}}_{\chi^2_{V^2}} + \underbrace{\sum\frac{({\rm CP}_{mod} - {\rm CP}_{data})^2}{\sigma^2_{\rm CP}}}_{\chi^2_{\rm CP}} \; .
\end{equation}
Here again, this is done for each binary model to obtain a $\chi^2(x,y,r)$ cube. The same procedure as in \citet{Absil11} is then repeated to assess the presence of a companion. This new method is more robust as we need to have a signature in both the CP and $V^2$ to detect a companion. We note however that the $V^2$ is also sensitive to centro-symmetric circumstellar emission, which creates a drop in visibility at all baselines \citep{DiFolco07}. Sometimes, the signature of a disc can be so strong that the combined $\chi^2$ is relatively high. To discriminate this kind of situation and identify bona fide point-like companions, we inspect the $\chi^2$ maps individually for the CP and the $V^2$. Indeed, if a companion is present, its signature will be seen in both maps (generally at the same position, except for marginal detections), while a disc will only show up in the $\chi^2_{V^2}$ map (assuming a symmetric disc). Furthermore, a small offset in the CP due to imperfect calibration could simulate a companion (false positive detection). In this case, the detection will generally not show up in the $\chi^2_{V^2}$ map. In some cases, the $\chi^2$ maps for both the CP and the $V^2$ show a significant detection, but not at the same position. These cases need to be investigated more carefully (e.g.\ by looking at secondary peaks in the $\chi^2$ maps) to draw definitive conclusions, when possible.

	\subsection{Defining the search region}

\onllongtab{
\small
\begin{longtable}{cccccccccc}
\caption{Results of the search for companion for the 92 stars observed within the \textsc{Exozodi} survey. The significance of the detection is given as a signal-to-noise ratio, expressed as a number of $\sigma$, for the combined $\chi^2$ and the two individual $\chi^2$ analyses. The best-fit angular separation, position angle and contrast are given, using both positive and negative contrasts as explained in Sect.~\ref{sub:detecriteria}. The last two columns give the upper limit on the contrast of companions in case of non detections, for completeness levels of 50\% and 90\% of the search region (except for the stars for which circumstellar emission was detected).} \label{tab:all}\\
\hline\hline
 & & & & & & & & Median & Percentile 90 \\
Star & Date & Signif. & Signif. & Signif. & Ang.\ separation & P.A. & Contrast & upper limit & upper limit \\
 & & (CP+$V^2$) & (CP) & ($V^2$) & (mas) & (deg) & (\%) & (\%) & (\%)\\
\hline
\endfirsthead
\caption{continued.}\\
\hline\hline
 & & & & & & & & Median & Percentile 90 \\
Star & Date & Signif. & Signif. & Signif. & Ang.\ separation & P.A. & Contrast & upper limit & upper limit \\
 & & (CP+$V^2$) & (CP) & ($V^2$) & (mas) & (deg) & (\%) & (\%) & (\%)\\
\hline
\endhead
\hline
\endfoot
\object{HD 142}    & 2012 Jul 26 & -5.44 & 4.24 & -5.39 & 67.50 & 157.80 & -1.16 & 0.8& 1.2\\
\object{HD 1581}   & 2012 Oct 15 & -3.15  & 2.88 & -2.51 & 77.28 & 102.33 & -0.58 & 0.7& 1.0\\
\object{HD 2262}   & 2012 Oct 16 & 3.90 & -2.96 & 5.00 & 13.95 & 104.53 & 0.81 & --- & --- \\
\object{HD 3302}   & 2012 Oct 17 & 1.21 & 1.55 & 1.28 & 12.98 & -74.36 & 0.41 & 0.8 & 1.1 \\
\object{HD 3823}   & 2012 Oct 15 & 2.15 & -3.07 & 1.75 & 12.35 & -111.37 &0.42 & 1.0 & 1.3 \\
\object{HD 4150}   & 2012 Dec 17 & 7.71 & 3.73  & 7.90 & 90.51 & 81.42 & 2.28 & --- & ---\\
\object{HD 7570}   & 2012 Jul 26 & 1.58 & -2.07 & -1.79 & 98.62 & 152.52 & -0.47 & 0.6 & 0.8\\
\object{HD 7788}   & 2012 Jul 24 & 1.87 & 18.57 & 10.23 & 1.58 & 18.43 & 1.78 & 2.2 & 2.8\\
\object{HD 10647}  & 2012 Oct 17 & -1.49 & -2.14 &  -1.25 & 83.16 & 129.14 & -0.45 & 0.7 & 0.9\\
\object{HD 11171}  & 2012 Oct 16 & 1.25 & 1.94 & 1.00 & 80.74 & -62.32 & 0.71 & 1.1 & 1.6\\
\object{HD 14412}  & 2012 Oct 15 & 2.73 & 2.03 & 3.85 & 30.70 & -59.68 & 7.37 & 1.2 & 1.7\\
\object{HD 15008}  & 2012 Jul 24 & 4.91 & -2.34 & 5.42 & 40.18 & 106.63 & 1.78 & ---  & --- \\
\object{HD 15798}  & 2012 Oct 16 & 6.25 & 1.98 & 14.01 & 29.47 & -104.74 &  1.76 & --- & --- \\
\object{HD 16555}  & 2012 Dec 18 & 105.32 & 26.11 & 212.70 & 78.69 & 40.88 & 51.26 & --- & --- \\
\object{HD 17051}  & 2012 Oct 17 & -1.51 & 1.78 & -1.76 & 66.54 & 87.85 & -0.33 &  0.6	& 0.8\\
\object{HD 17925}  & 2012 Oct 15 & -2.04 & -3.57 & 1.44 & 55.43 & 111.71 & -0.32 &  0.7& 0.9\\
\object{HD 19107}  & 2012 Oct 15 & 2.81 & 2.50 & 3.21 & 23.16 & 122.66 & 0.56 &0.8 & 1.2\\
\object{HD 20766}  & 2012 Oct 15 & -1.44 & -2.11 & 0.95 & 81.90 & -71.12 & -0.39 & 0.6 & 0.8\\
\object{HD 20794}  & 2012 Oct 16 &2.15 & 4.53 & 2.05 & 33.59 & 94.27 & 3.62 & --- & ---\\
         $\cdots$  & 2012 Dec 17 & 3.96 & -1.76 & 5.96 & 30.21 & -24.44 & 1.15 & --- & ---\\
\object{HD 20807}  & 2012 Oct 16 & -2.58 & -2.59 & -3.02 & 22.64 & 59.47 & -1.65 & 1.1 & 1.7\\
\object{HD 22001}  & 2012 Oct 17 & 1.86 & -2.14 & 1.93 & 85.29 & 123.04 & 0.35 & 0.7 & 0.9\\
\object{HD 23249}  & 2012 Oct 16 & 2.93 & 2.70 & 5.86 & 2.55 & 78.69 & 12.86 & --- & ---\\
         $\cdots$  & 2012 Dec 17 & 4.80 & -4.96 & 7.84 & 3.81 & 156.80 & 5.46 & --- & ---\\
\object{HD 25457}  & 2012 Oct 15 & -1.28 & 1.13 & -1.21 & 127.30 & 134.04 & -0.45 & 0.4 & 0.6\\
\object{HD 28355}  & 2012 Dec 16 & 5.23 & 2.44 & 7.94 & 96.82 & 116.70 & 0.85 & --- & ---\\
\object{HD 29388}  & 2012 Dec 16 & 111.86 & -53.65 & 130.18  & 11.07& 71.56 & 3.01 & --- & ---\\
\object{HD 30495}  & 2012 Oct 15 & -1.34 & -2.41 & -1.68 & 36.75 & 104.98 & -0.27 & 0.5 & 0.8\\
\object{HD 31295}  & 2012 Dec 16 & 1.73 & 2.41 & 1.29 & 72.68 & -154.32 & 0.29 & 0.9 & 0.6\\
\object{HD 31925}  & 2012 Oct 15 & 2.03 & -3.42 & 1.76 & 103.21 & 66.29 & 0.37 & 1.0 & 1.6\\
         $\cdots$  & 2012 Oct 17 & 1.80 & 1.43 & 1.84 & 58.52 & 91.47 & 0.58 & 1.1 & 1.6\\
\object{HD 33111}  & 2012 Dec 16 &5.15 & 2.73 & 5.99 & 44.52 & 124.94 & 0.76 & 1.5 & 2.0\\
         $\cdots$  & 2012 Dec 18 & 1.11 & 1.54 & 1.01 & 113.24 & 137.50 & 0.68 & 0.8 & 1.2\\
\object{HD 33262}  & 2012 Oct 17 & 2.93 & 3.51 & 1.51 & 48.93 & -172.37 & 0.32 & 0.6 & 0.9\\
\object{HD 34721}  & 2012 Oct 17 & -2.41 & 2.33 & -2.10 &  93.40 & 137.17 & -0.61 & 0.6 & 0.8\\
\object{HD 38858}  & 2012 Oct 17 & -2.59 & 2.05 & -2.53 & 83.93 & -54.70 & -0.69 & 0.7 & 1.0\\
\object{HD 39060}  & 2012 Oct 17 & 5.79 & -2.34 & 7.83 & 91.96 & 122.57 & 1.30 & --- & ---\\
\object{HD 40307}  & 2012 Dec 18 & -1.55 & -1.62 & -1.48 & 96.48 & -49.62 & -0.54 & 0.8 & 1.0\\
\object{HD 43162}  & 2012 Dec 18 & 2.20 & 2.32 & 2.62 & 74.61 & -30.17 & 0.45 & 0.9 & 1.2\\
\object{HD 45184}  & 2012 Dec 16 & 2.08 & 2.44 & 2.59 & 10.61 & 45.00 & 0.42 & 0.9 & 1.2\\
\object{HD 53705}  & 2012 Dec 17 & 1.72 & 2.56 & 1.89 & 90.51 & 8.58 & 0.47 & 0.8 & 1.1\\
\object{HD 56537}  & 2012 Dec 17 & -3.84 & 4.64 & -3.25 & 4.53 & 6.34 & -0.94 & 0.5 & 0.9\\
\object{HD 69830}  & 2012 Dec 17 & 2.48 & 3.53 & 1.57 & 59.35 & -66.67 & 0.51 & 0.7 & 1.0 \\
\object{HD 71155}  & 2012 Dec 16 & -1.78 & -3.00 & 1.19 & 35.67 & -95.63 & -0.31 & 0.7 & 1.1\\
\object{HD 72673}  & 2012 Dec 18 & 1.29 & -1.12 & 1.31 & 5.15 & 119.05 & 2.7 & 0.7 & 0.9\\
\object{HD 76151}  & 2012 Dec 18 & 1.80 & -1.80 & 1.43 & 71.78 & -63.97 & 0.60 & 0.8 & 1.1\\
\object{HD 76932}  & 2012 Dec 17 & -1.14 & 1.56 & -2.54 & 26.09 & -77.83 & -0.95 & 0.6 & 0.9\\
\object{HD 82434}  & 2012 Dec 17 &1.69 & 2.56 & 1.51 & 62.93& -70.99& 0.79 & 1.2 & 1.9\\
\object{HD 88955}  & 2012 Dec 18 & -1.67 & -2.07 & -1.93 & 64.73 & 113.20 & -0.76 & 0.5 & 0.8\\
\object{HD 90132}  & 2012 Apr 29 & -1.97 & -2.32 & -1.73 & 88.57 & 48.66 & 0.80 & 1.3 & 1.8\\
\object{HD 91324}  & 2012 Apr 29 & 1.31 & 2.23 & 1.14 & 97.10 & 70.44 & 0.36 & 0.7 & 0.9\\
\object{HD 99211}  & 2012 Apr 28 & 1.97 & 2.69 & 2.17 & 96.17 & -44.58 & 0.42 & 0.9 & 1.2\\
\object{HD 102365} & 2012 Apr 29 & 2.02 & 2.93 & 2.28 & 42.48& 42.14 & 0.57 & 1.1 & 1.5\\
\object{HD 104731} & 2012 Apr 30 & 2.69 & -1.98 & 3.44 & 1.58 & 18.43 & 4.73 & 0.9 & 1.2\\
\object{HD 108767} & 2012 Apr 30 & 3.08 & 1.70 & 3.64 & 57.55 & 87.51 & 0.51 & --- & ---\\
\object{HD 109787} & 2012 Apr 30 & -1.96 & -2.38 & -1.89 & 65.88 & 147.39 & -0.43 & 0.6 & 0.8\\
\object{HD 115617} & 2012 Apr 29 & -1.15 & -2.17 & -1.01 & 99.26 & 112.82 & -0.27 & 0.6 & 0.8\\
\object{HD 120136} & 2012 Apr 30 & -1.39 & -2.31 & -1.21 & 82.93 & 174.12 & -0.34 & 0.6 & 0.9\\
\object{HD 128898} & 2012 Apr 29 & 1.30 & 2.31 & 1.14 &43.23 & -60.17 & 0.42 & 0.7 & 0.9\\
\object{HD 129502} & 2012 Apr 29 & -1.49 & -2.17 & -1.16 & 67.41 & 139.81 & -0.28 & 0.4 & 0.6\\
\object{HD 130109} & 2012 Jul 25 & -1.63 & 1.72 & -1.64 & 80.62 & -110.08 & -0.51 & 0.9 & 1.4\\
\object{HD 134083} & 2012 Jul 24 & -2.12 & -3.30 & -1.99 & 132.98 & -46.52 & -0.71 & 1.0 & 1.6\\
\object{HD 135379} & 2012 Apr 29 & 1.18 & 1.35 & 1.26 & 46.82 & -83.25 & 0.57 & 1.4 & 1.7\\
\object{HD 136202} & 2012 Jul 24 & -1.93 & -1.91 & -2.47 & 21.51 & -88.67 & -0.82 & 0.7 & 1.2\\
\object{HD 139664} & 2012 Apr 29 & 2.17 & 3.43 & 1.25 & 121.12 & 127.96 & 0.61 & 0.8 & 1.2\\
\object{HD 141891} & 2012 Apr 29 & -1.48 & -2.33 & -1.29 & 102.82 & 106.67 & -0.55 & 0.7 & 1.0\\
\object{HD 149661} & 2012 Apr 29 & -2.89 & -2.52 & -3.44 & 118.22 & 42.26 & -2.70 & 2.9 & 4.9\\
          $\cdots$ & 2012 Apr 30 & 1.66 & 1.98 & 2.05 & 48.71 & 162.68 & 2.08 & 0.6 & 1.0\\
\object{HD 152391} & 2012 Apr 30 & 1.36 & 1.84 & 1.49 & 37.39 & 119.65 & 1.00 & 0.7 & 1.1\\
\object{HD 160032} & 2012 Apr 30 & -1.09& -2.01 & -0.95 & 60.84 & 6.13 & -0.16 & 0.3 & 0.5\\
\object{HD 160915} & 2012 Apr 29 & -0.98 & 2.06 & -0.89 & 53.20 & -49.57 & -0.36 & 1.3 & 1.8\\
         $\cdots$ & 2012 Apr 30 & 2.18 & -4.06 & 1.63 & 91.36& -97.86 & 0.31 & 0.5 & 1.0\\
\object{HD 164259} & 2012 Jul 26 & -1.63 & 1.73 & -1.64 & 80.62 & 176.80 & -0.45 & 0.5 & 0.7\\
\object{HD 165777} & 2012 Jul 26 & 2.58 & -1.70 & 2.76 & 95.73& 67.59 & 0.98 & 1.2 & 1.6\\
\object{HD 172555} & 2012 Jul 25 & 2.68 & -2.94 & 3.53 & 79.97 & 158.27 & 0.61 & 1.1 & 1.5\\
\object{HD 178253} & 2012 Jul 24 & 2.37 & -4.19 & 1.91 & 23.03 & -27.12 & 0.48 & 0.9 & 1.4\\
\object{HD 182572} & 2012 Jul 25 & 1.10 & 1.58 & 1.34 & 103.32 & 123.15 & 0.16 & 0.5 & 0.6\\
\object{HD 188228} & 2012 Jul 26 & 2.21 & 2.14 & 4.32 & 67.55 & -92.12 & 0.68 & 1.1 & 1.5\\
\object{HD 192425} & 2012 Jul 25 & -1.76 & 1.57 & -3.29 & 28.26 & 103.30 & -0.95 & 0.6 & 0.8\\
\object{HD 195627} & 2012 Jul 24 & 2.14 & -3.69 & -2.21 & 56.20 & 156.40 & 0.51 & 1.0 & 1.5\\
\object{HD 197157} & 2012 Jul 26 & 1.93 & -2.37 & 2.87 &81.62 & -76.90 & 0.61 & 1.0 & 1.3\\
\object{HD 197692} & 2012 Apr 30 & -1.67 & -1.82 & -2.04 & 10.12 & -147.09 & -0.35 & 0.4 & 0.6\\
\object{HD 202730} & 2012 Jul 24 & 12.25 & 6.69 & 21.31 & 61.74 & -21.37 & 87.44 & --- & ---\\
\object{HD 203608} & 2012 Jul 24 & -2.79 & 3.05 & -3.82 & 46.76 & 65.35 & N/A & 0.5 & 0.7\\
\object{HD 206860} & 2012 Jul 25 & 2.19 & -1.37 & 2.76 & 36.44 & -63.08 & 1.16 & 0.9 & 1.4\\
\object{HD 207129} & 2012 Jul 26 & 1.41 & 2.28 & 1.51 & 95.04 & -25.89 & 0.48 & 0.5 & 0.7\\
\object{HD 210049} & 2012 Jul 25 & 1.95 & 3.92 & 1.55 & 82.39 & -50.42 & 0.82 & 1.1 & 1.6\\
\object{HD 210277} & 2012 Oct 15 & -2.00 & -3.38 & -2.05 & 20.89 & 68.96 & -0.94 & 0.9 & 1.3\\
\object{HD 210302} & 2012 Jul 25 & 2.60 & -3.13 & 3.82 & 41.38 & 115.01 & 0.57 & 1.0 & 1.5\\
\object{HD 210418} & 2012 Jul 26 & -3.43 & 4.78 & -3.83 & 3.53 & 81.87 & -2.17 & 0.7 & 1.0\\
\object{HD 213845} & 2012 Jul 25 & -1.86 & 2.11 & -1.59 & 86.09 & 115.82 & -0.45 & 0.6 & 0.8\\
\object{HD 214953} & 2012 Oct 17 & -1.55 & -1.83 & -1.37 & 52.56 & -27.78 & -0.49 & 0.6 & 0.8\\
\object{HD 215648} & 2012 Oct 14 & -2.15 & 3.03 & -4.82 & 3.81 & 66.80 & -1.08 & 0.4 & 0.7\\
\object{HD 215789} & 2012 Jul 24 & -2.90 & -3.61 & -2.18 & 37.63 & 109.40 & -0.63 & 0.5 & 0.7\\
\object{HD 216435} & 2012 Oct 17 & -1.58 & 1.64 & -2.02 & 20.70 & -127.14 & -0.54 & 0.7 & 0.9\\
\object{HD 219482} & 2012 Jul 25 & 2.18 & 4.50 & 1.58 & 111.95 & 144.82 & 0.51 & 0.7 & 0.9\\
\object{HD 219571} & 2012 Jul 26 & 1.66 & -2.11 & 1.61 & 53.39 & 123.54 & 0.38 & 0.7 & 0.9\\
\object{HD 224392} & 2012 Jul 26  & 15.06 & -20.44 & 7.69 & 16.81 & -120.38 & 1.31 & --- & ---\\
\end{longtable}
}

\begin{table*}[t]
\caption{Summary of the stars showing a significance level higher than $3\sigma$ based on the analysis of the combined $\chi^2$ (CP+$V^2$). The significance of the detection based on the separate analysis of the CP and the $V^2$ is also given, and used to classify the nature of the detected excess between disc (no significant CP signature) and companion (significant CP signature).}
\label{tab:detections}
\centering
\begin{tabular}{cccccc}
\hline \hline
Name & Date & Significance & Significance & Significance & Nature \\
 &  & (CP+$V^2$) & (CP only) & ($V^2$ only) &  \\
\hline
HD~2262 & 2012 Oct 16 & 3.9 & 2.9 & 5.0 & disc \\
HD~4150 & 2012 Dec 17 & 7.7 & 3.7& 7.9 & companion \\
HD~15008 & 2012 Jul 24 & 4.9 & 2.3 & 5.4 & disc \\
HD~15798 & 2012 Oct 16 & 6.3& 2.0 & 14.0 & (see Sect.~\ref{sub:results}) \\
HD~16555 & 2012 Dec 18 & 105.3 & 26.1 & 212.7 & companion \\
HD~20794 & 2012 Oct 15 & 2.2 & 4.5 & 2.1 & disc \\
$\cdots$ & 2012 Dec 17 & 4.0 &1.8 & 6.0 & $\cdots$ \\
HD~23249& 2012 Oct 15 & 2.9 & 2.7 & 5.9 & disc \\
$\cdots$ & 2012 Dec 16 & 4.8 & 5.0 & 7.8 & $\cdots$ \\
HD~28355 & 2012 Dec 15 & 5.2 & 2.4 & 7.9 & disc \\
HD~29388 & 2012 Dec 16 & 111.7 & 53.6 & 130.1 &companion  \\
HD~39060 & 2012 Oct 16 &  5.8 & 2.3 & 7.8 & disc \\
HD~108767 & 2012 Apr 30 & 3.1 & 1.7 & 3.6 & disc \\
HD~202730 & 2012 Jul 24 & 12.2 & 6.7 & 21.3 & companion  \\
HD~224392 & 2012 Jul 26 & 15.1 & 20.4 & 7.7 & companion \\
\hline
\end{tabular}
\end{table*}

There are two main limitations to the detection of faint companions with interferometry: the dynamic range in the observations, and the limited field of view. Before starting our search for companions, we first need to define a suitable search region. This region is limited, as explained in \citet{Absil11}, by three factors. The first relates to the beams being injected into single-mode fibres for modal filtering and beam combination. As explained in \citet{Absil11}, we can consider that the PIONIER fibres have a Gaussian transmission profile with a full width at half maximum of 420~mas under typical turbulence conditions at Paranal. Second, the fringe packets associated with the primary star and its companion need to be within the OPD scan length and to partially overlap in order to contribute to the coherent signal measured by the PIONIER data reduction pipeline. The shortest scans performed by PIONIER are about $60~\mu$m long, and we estimate the size of one fringe packet to be $\lambda^2/\Delta\lambda \simeq 27\mu$m when the H band signal is dispersed onto three spectral channels. The maximum OPD separation ($\Delta$OPD) between the two packets can therefore not exceed $27~\mu$m. This separation is related to the angular separation $\Delta\theta$ of the two objects in the sky: $\Delta {\rm OPD} = B \Delta\theta \cos \alpha$, with $B \cos \alpha$ the projected baseline. Considering a maximum baseline $B = 36$~m, we find a maximum angular separation $\Delta\theta_{\rm max} \simeq 150$ mas.

The last limitation to the field of view comes from the need to properly sample the closure phase variations as a function of wavelength: companions located too far away can create aliasing inside the search region if their closure phase signal has a period shorter than four times the spectral channel size (for Nyquist sampling). This phenomenon can be partly mitigated by repeating the observations in time. Here, we only have three consecutive OBs for most of our targets. We will therefore compute a pessimistic non-ambiguous field of view as if we only had one sample in time for each star. As already detailed by \citet{Absil11}, the period in the closure phase signal is roughly given by\footnote{This formula can be obtained if we assume that the periodicity in the $V^2$ and in the CP are the same. Using a mean baseline $B$, we can then determine $\Delta\lambda$ using the $2\pi$ periodicity of the $\sin(2\pi B\theta/\lambda)$ term in the $V^2$ formula, i.e.\ $2\pi B\theta/\lambda= 2\pi B\theta /(\lambda +\Delta \lambda) + 2 \pi$.} $P_{\lambda} \simeq \lambda^2/(B\Delta\theta -\lambda)$, and must be larger than four times the spectral channel size ($0.1~\mu$m here). This lead to $\Delta\theta_{\rm max} \simeq 87$~mas for our mean baseline $B \simeq 20$~m. We will therefore consider a search region of about 100~mas in this study. It must be noted that companions located up to 150~mas can nonetheless have a significant signature in our data.

	\subsection{Defining the detection criterion} \label{sub:detecriteria}

\begin{figure}[t]
\begin{center}
\includegraphics[scale=0.5]{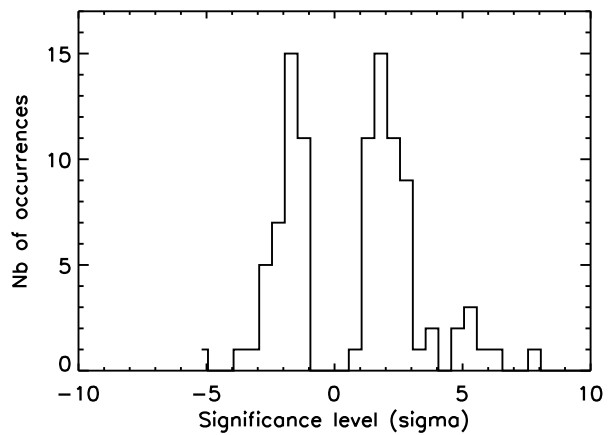}
\caption{Statistics of the (signed) significance level for the 92 stars based on the combined $\chi^2$, taking into account the CP and $V^2$. Five stars with a significance level higher than $10\sigma$ are not represented here for the sake of clarity.}
\label{fig:histosignif}
\end{center}
\end{figure}

\begin{figure*}[!t]
\centering
\includegraphics[scale=0.33]{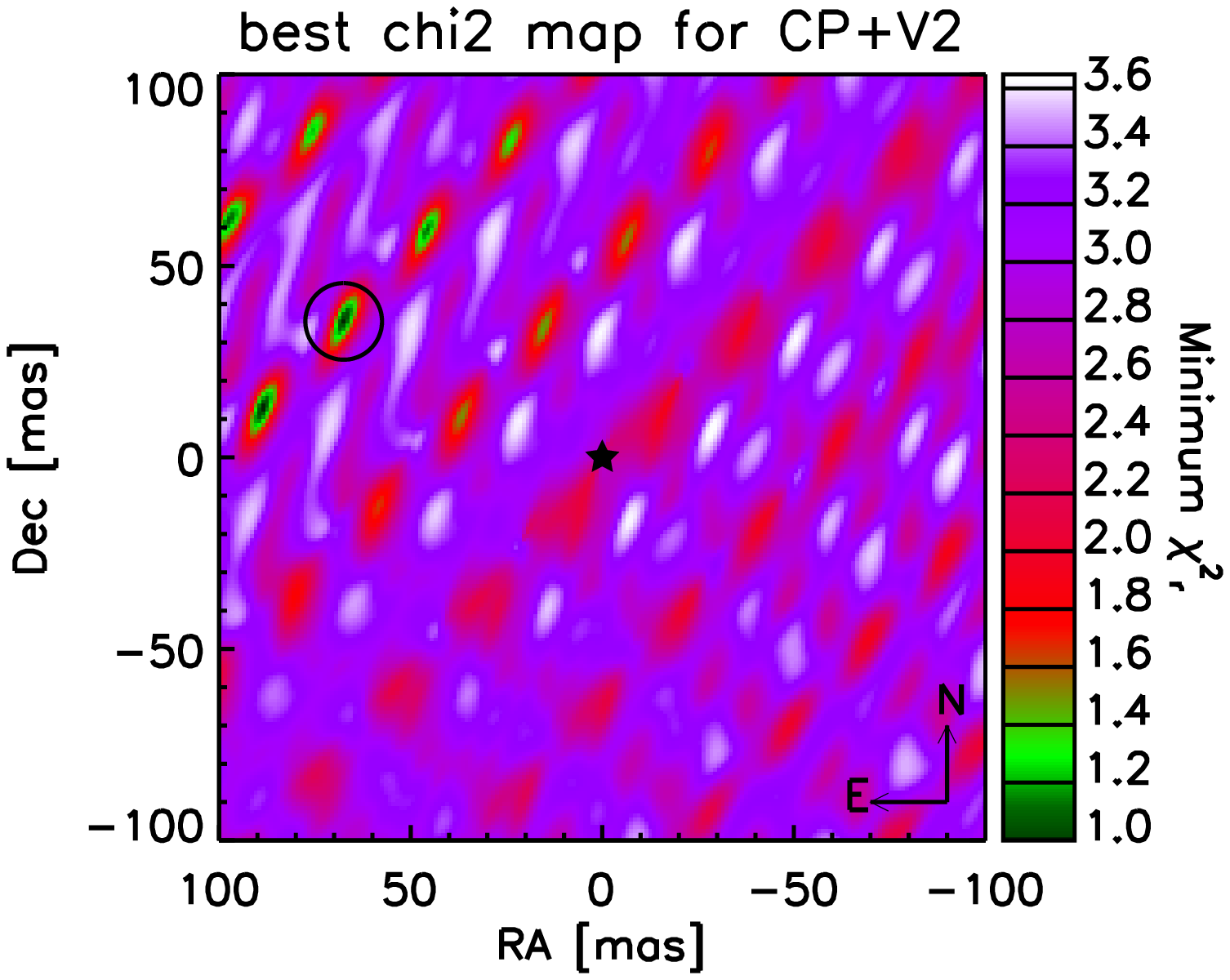}
\includegraphics[scale=0.33]{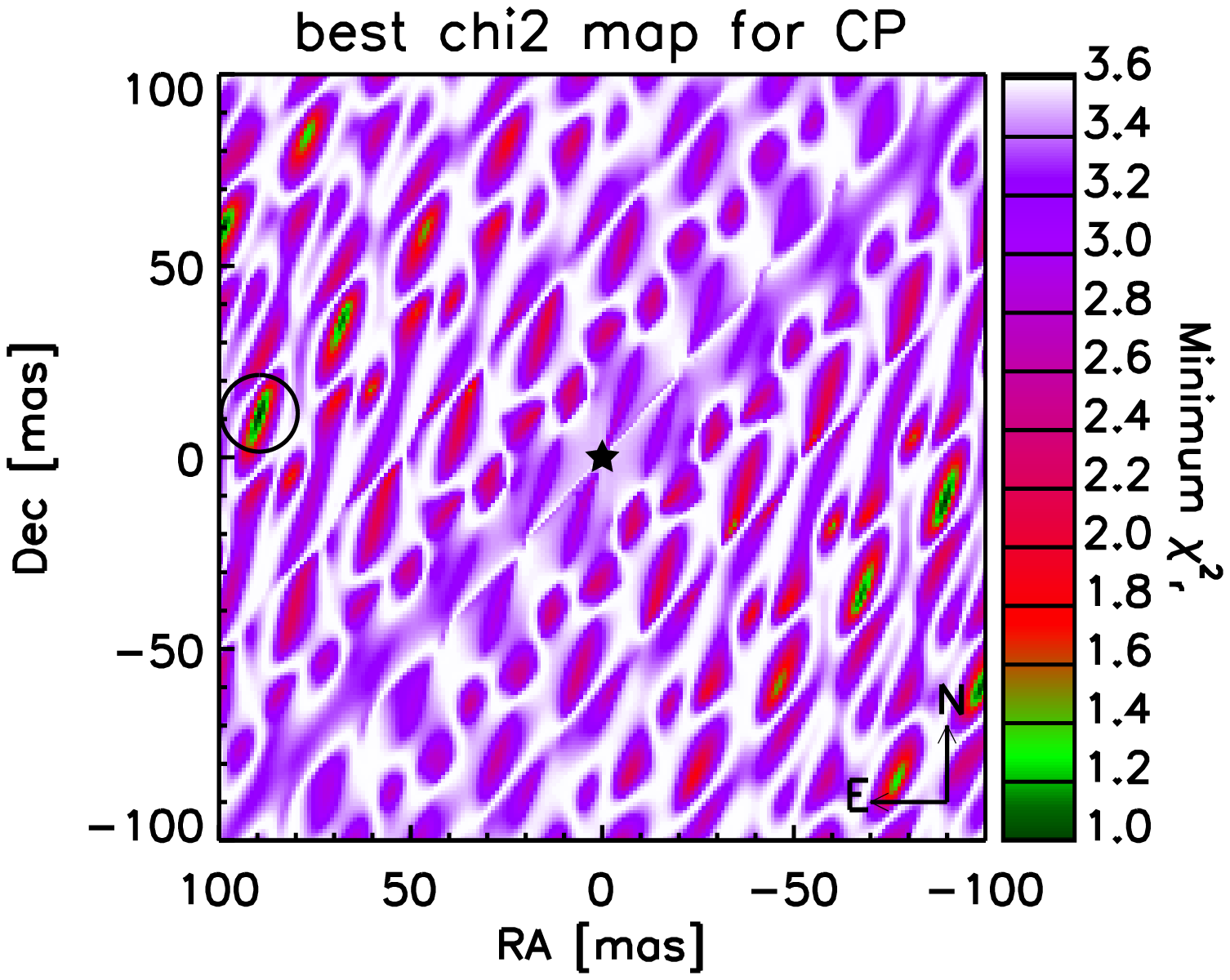}
\includegraphics[scale=0.33]{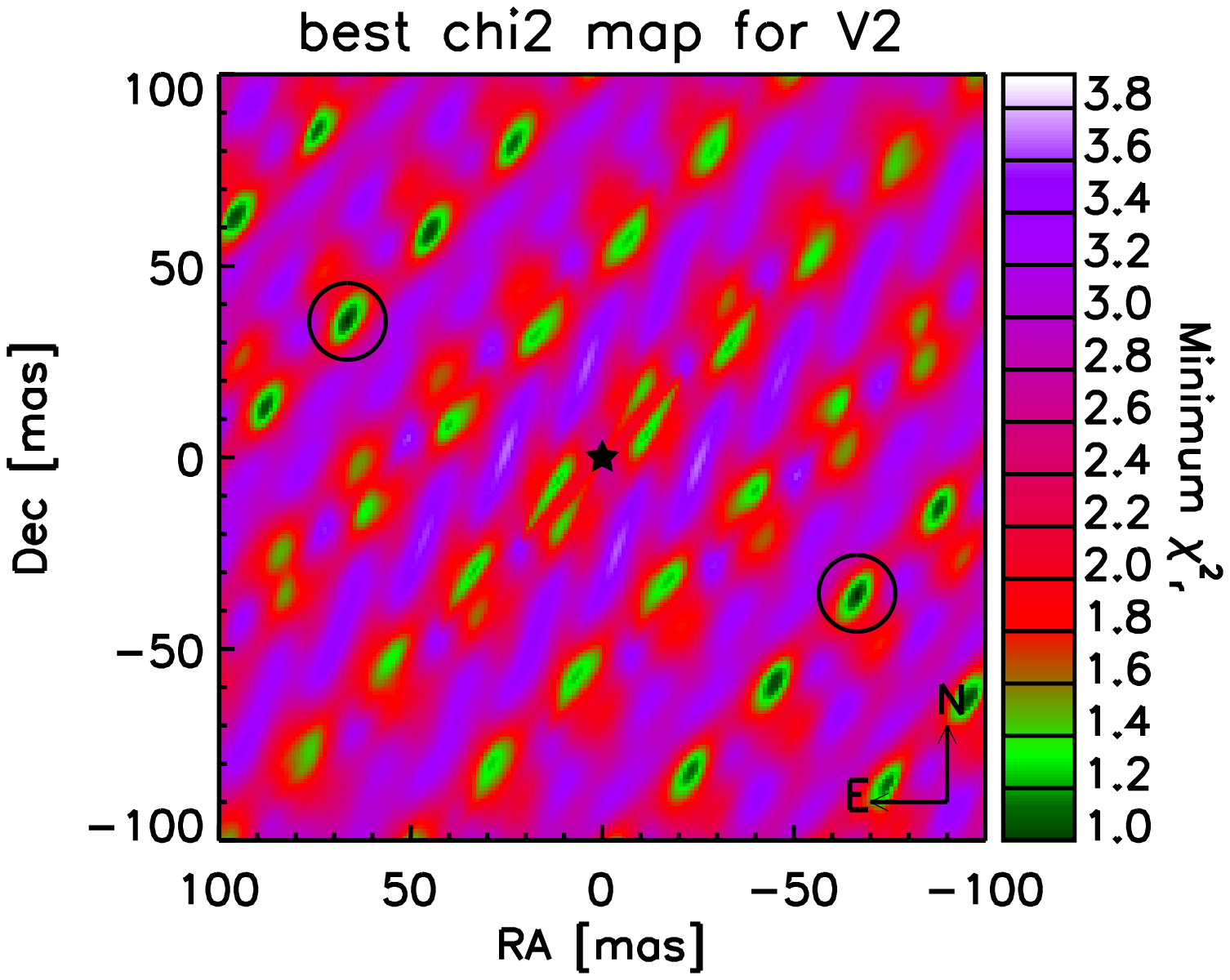}\\
\includegraphics[scale=0.33]{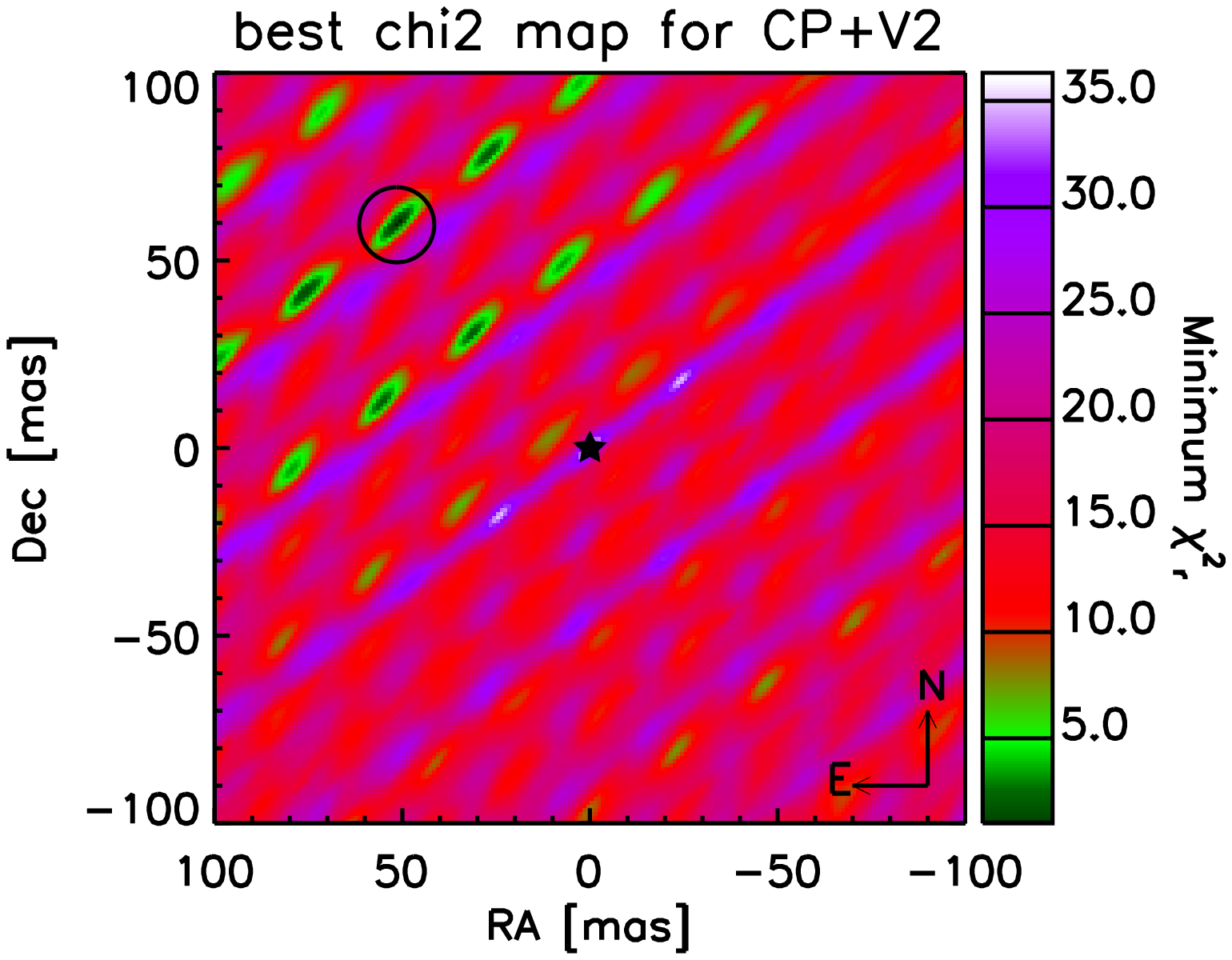}
\includegraphics[scale=0.33]{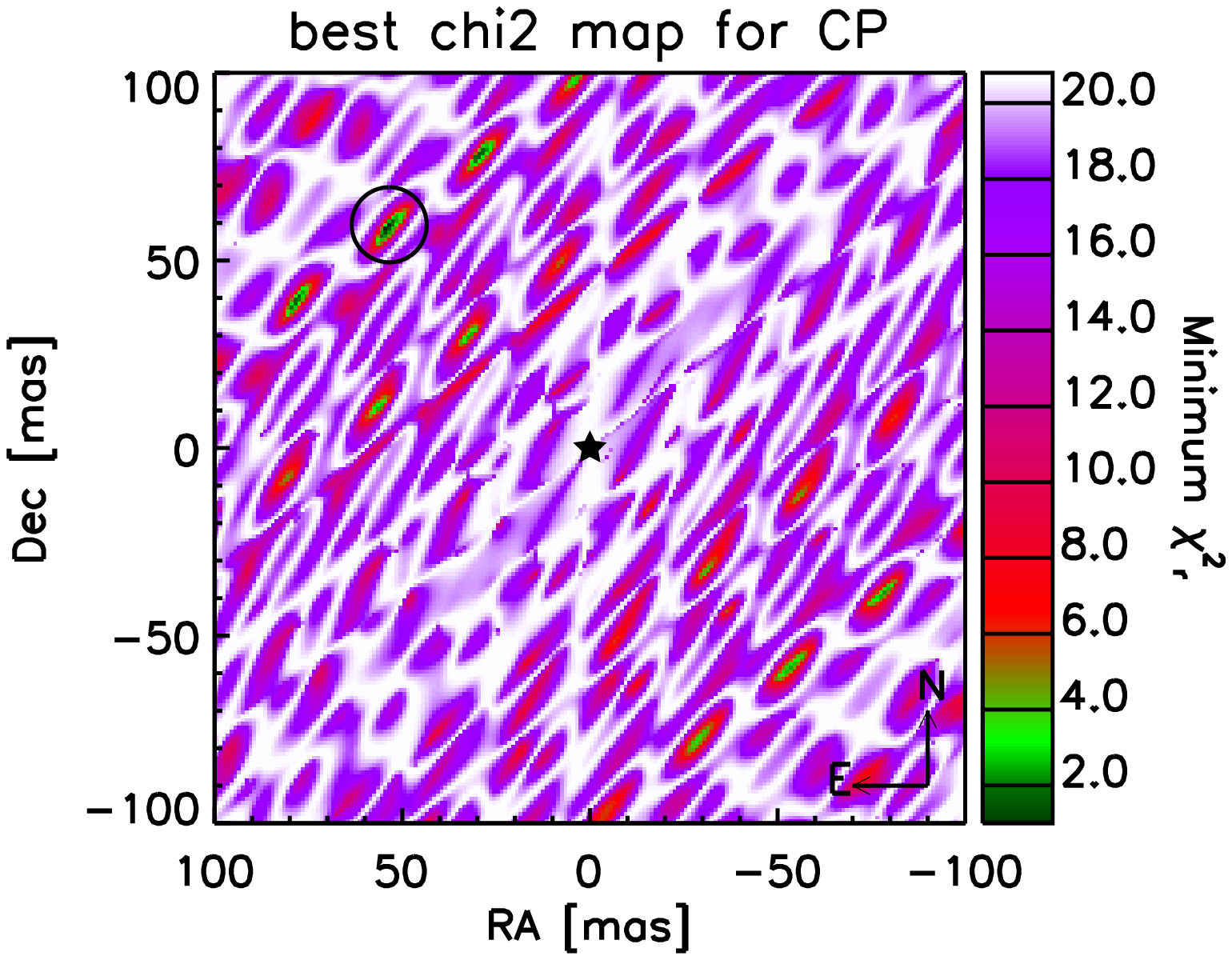}
\includegraphics[scale=0.33]{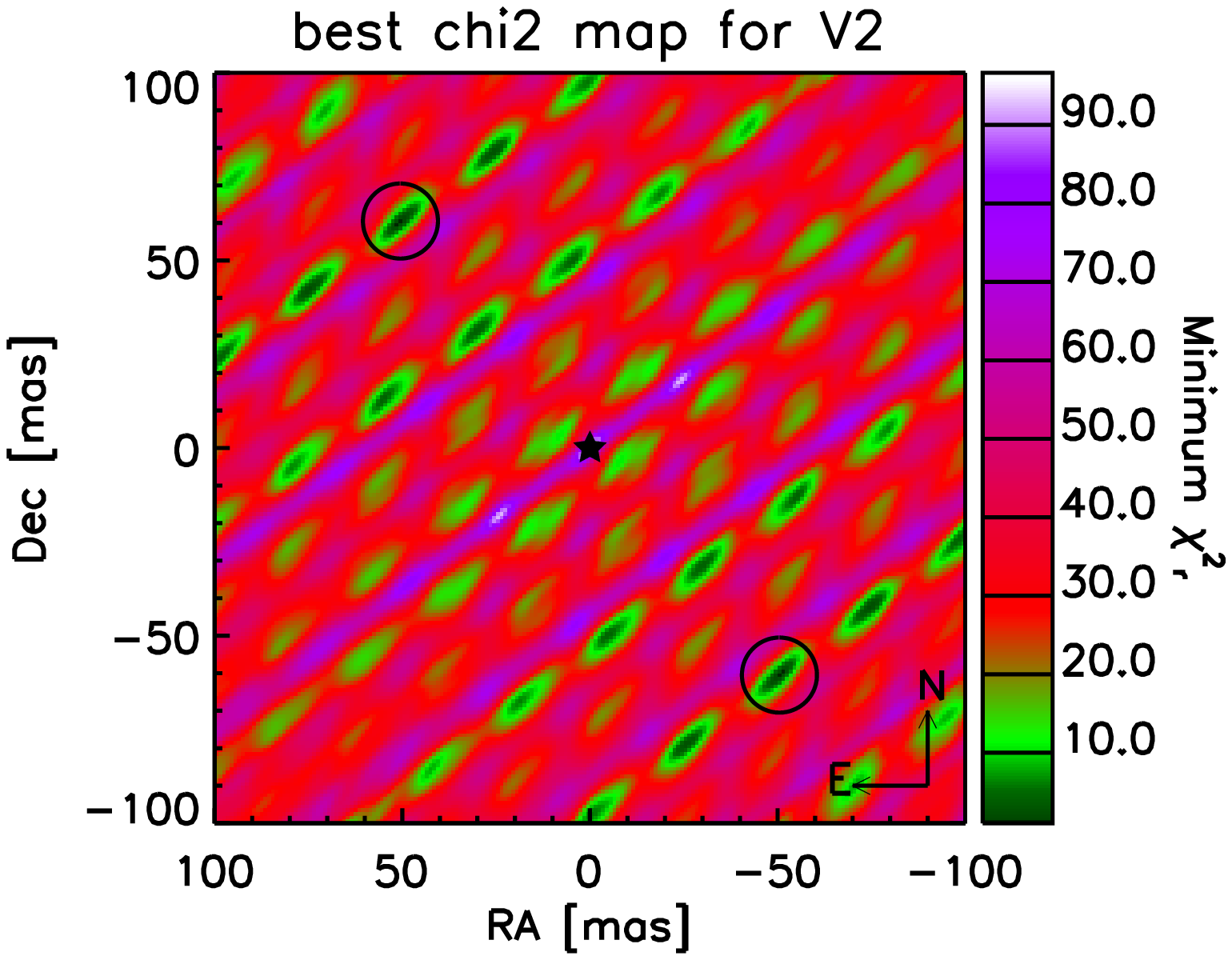}\\
\includegraphics[scale=0.33]{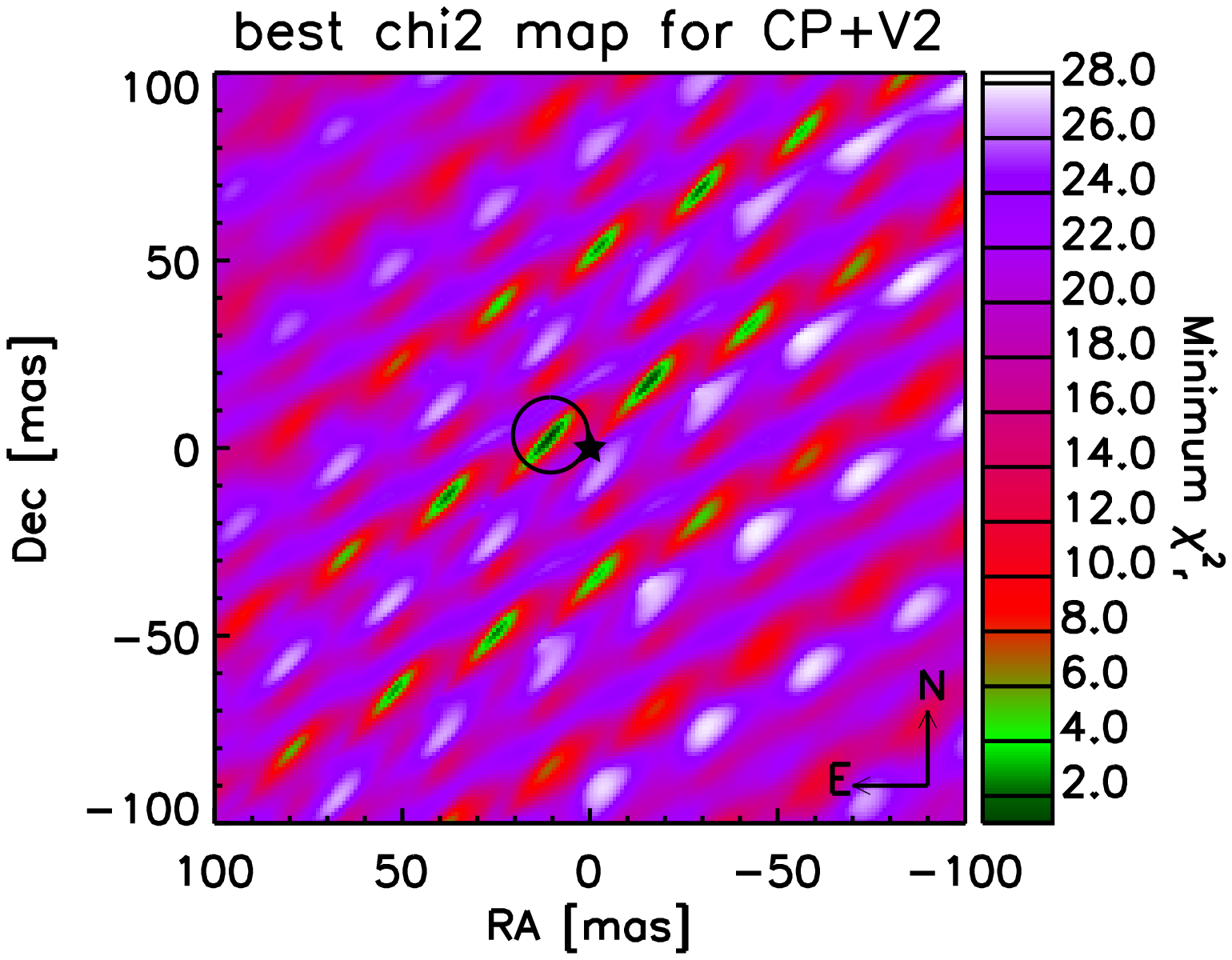}
\includegraphics[scale=0.33]{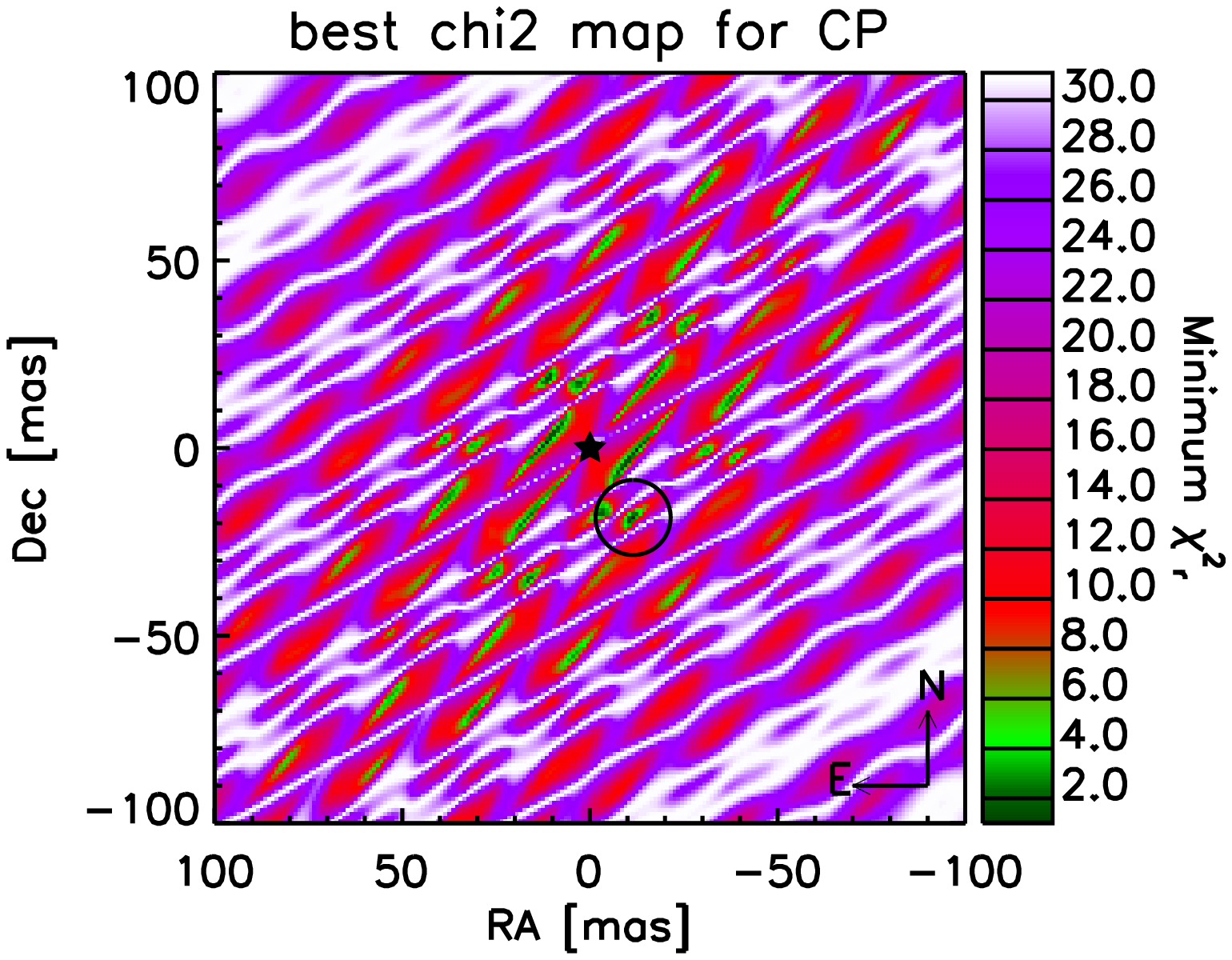}
\includegraphics[scale=0.33]{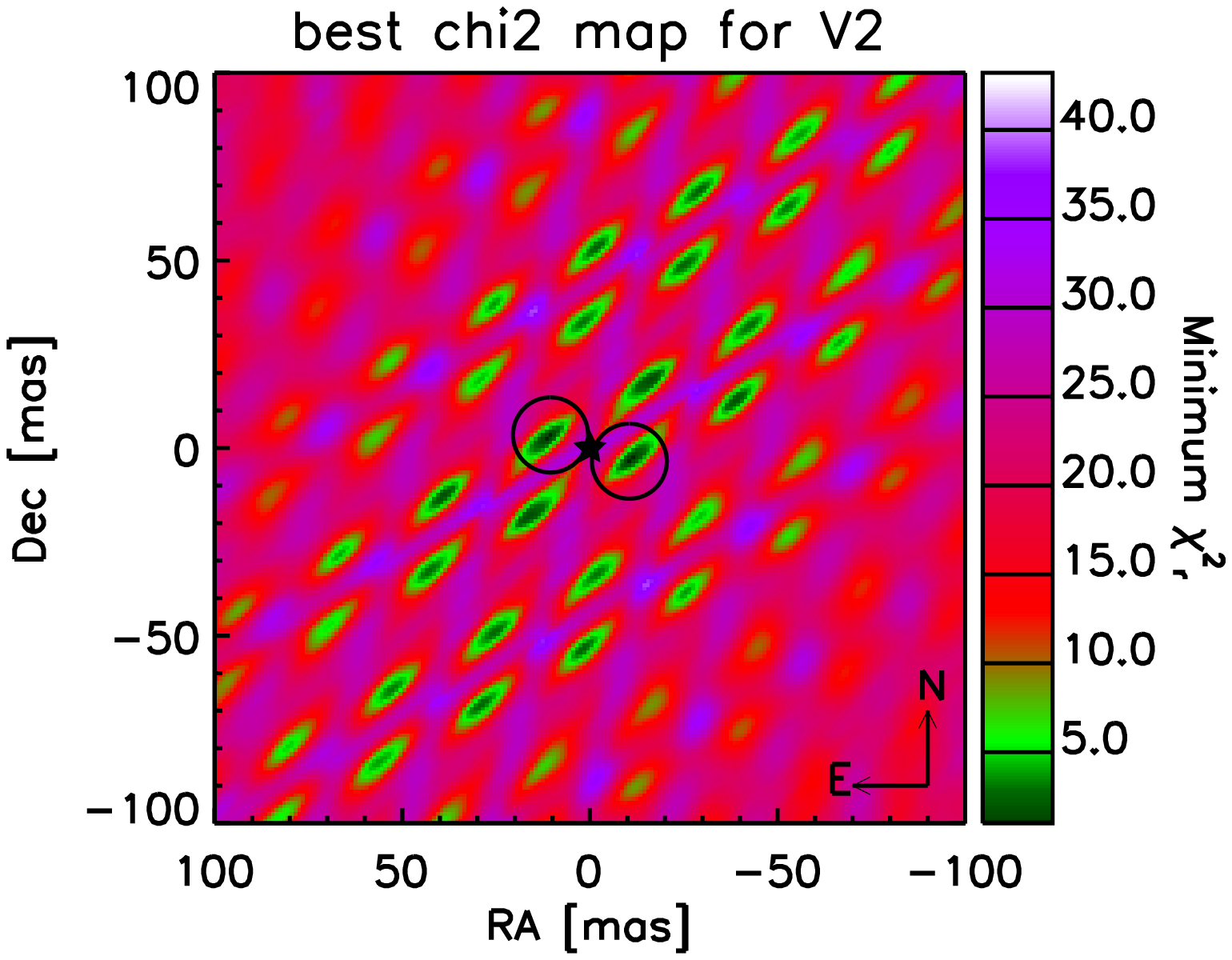}\\
\includegraphics[scale=0.33]{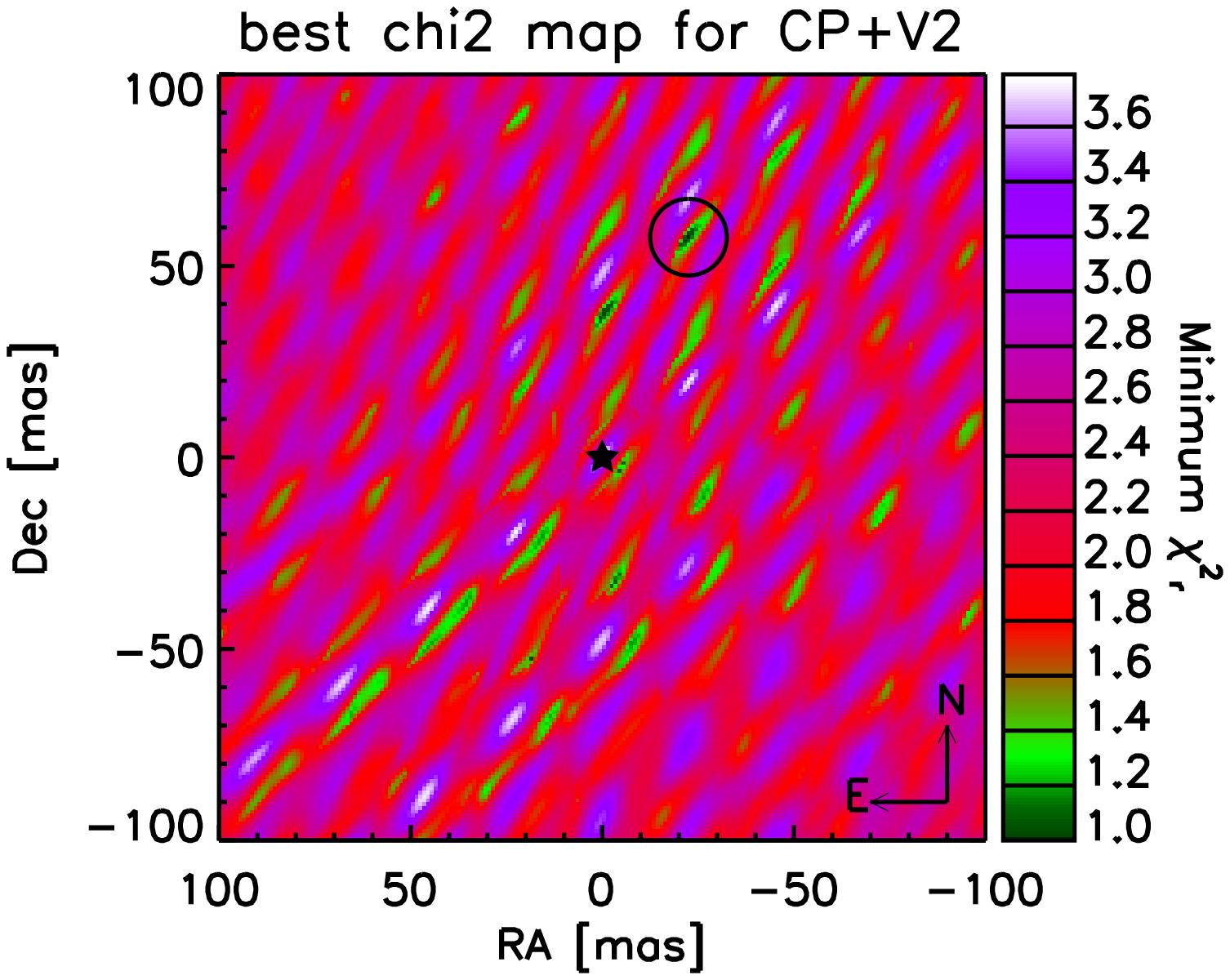}
\includegraphics[scale=0.33]{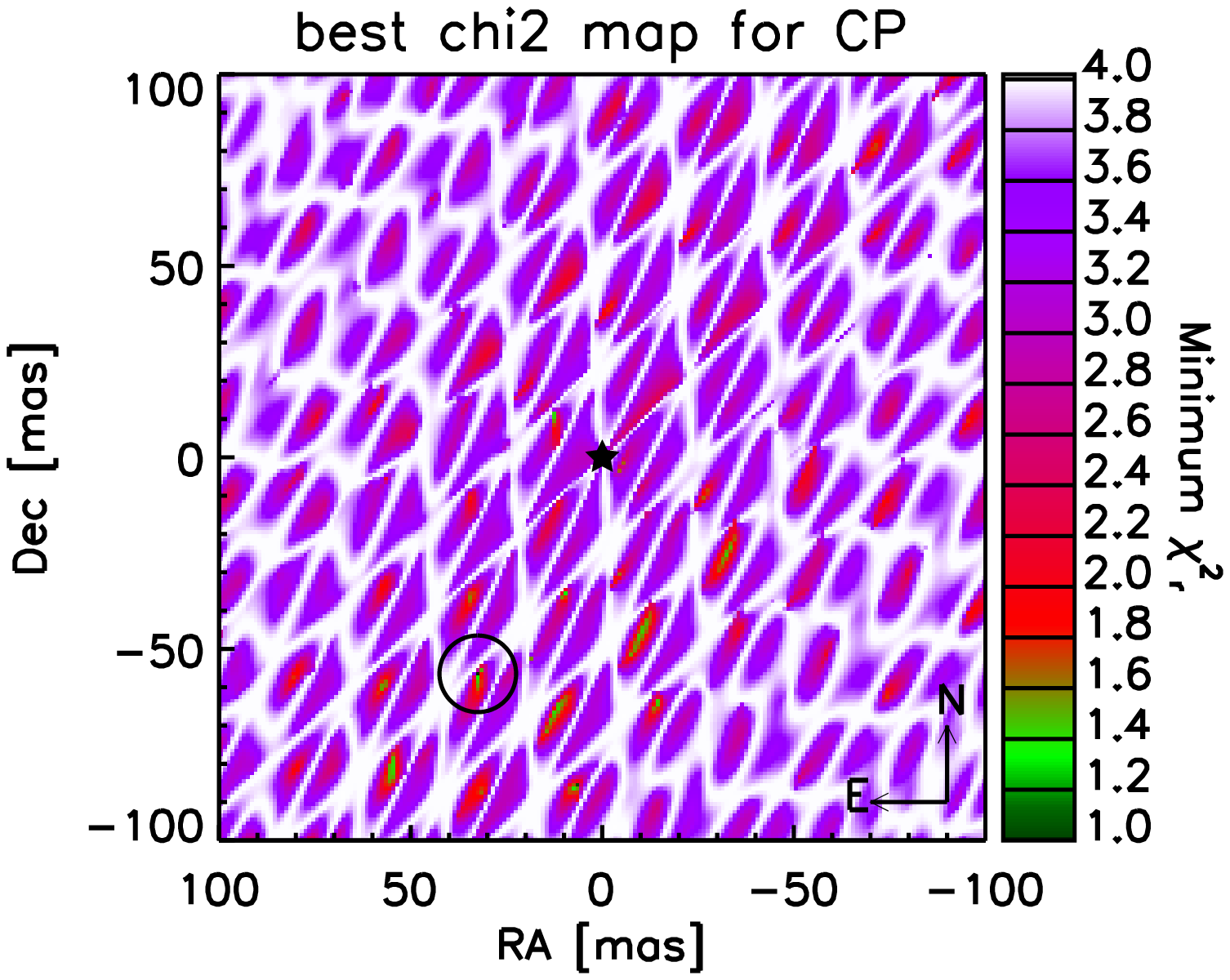}
\includegraphics[scale=0.33]{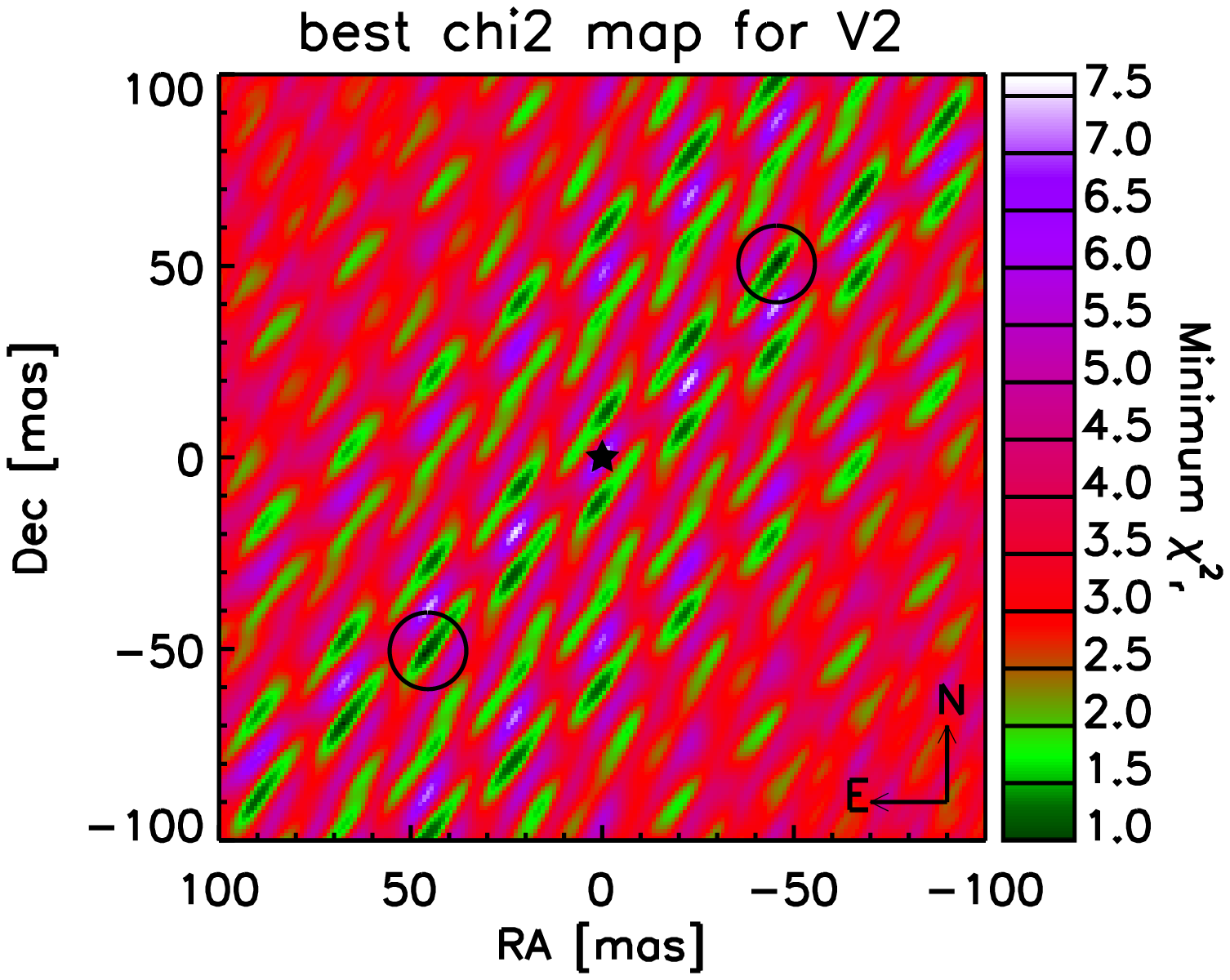}\\
\includegraphics[scale=0.33]{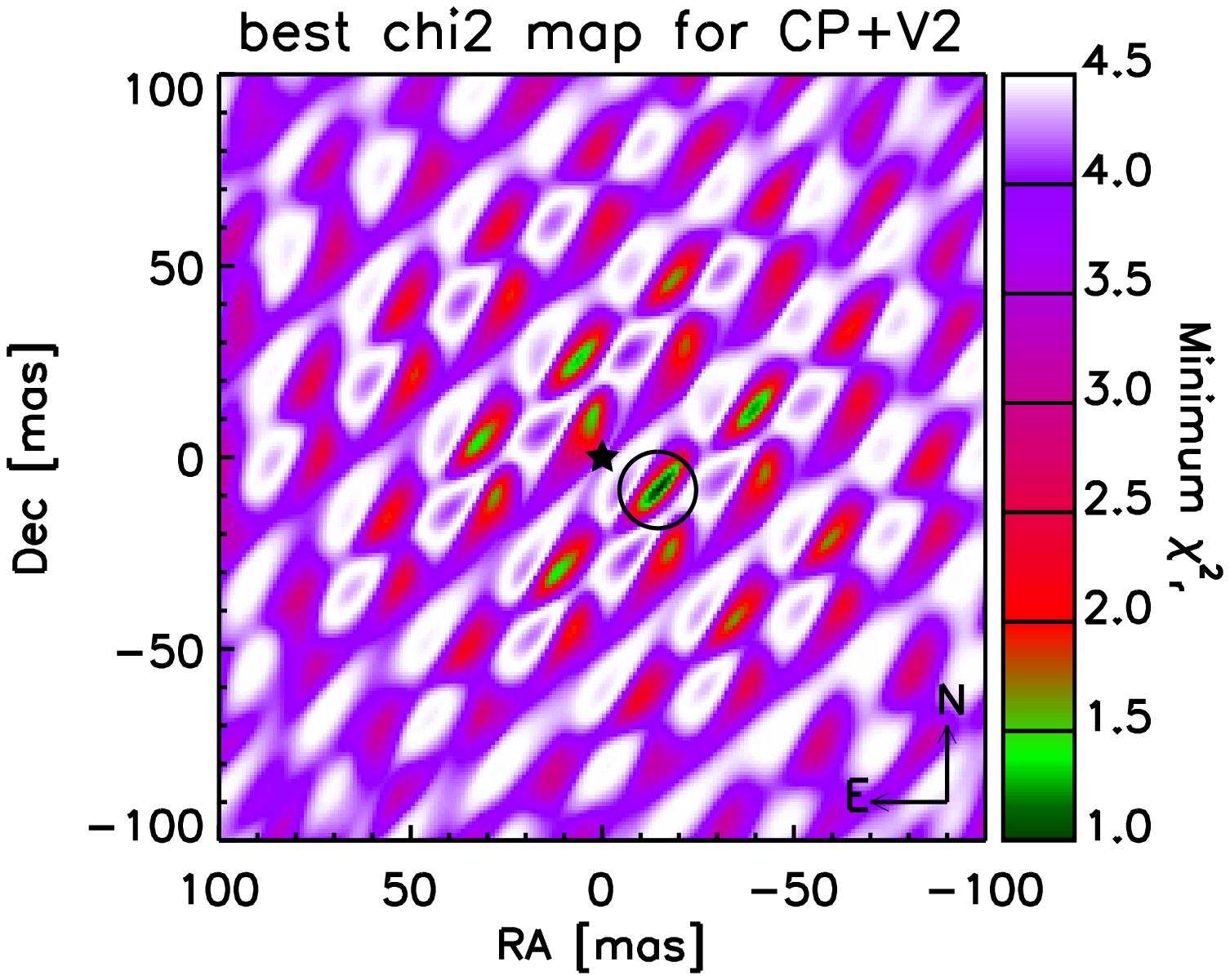}
\includegraphics[scale=0.33]{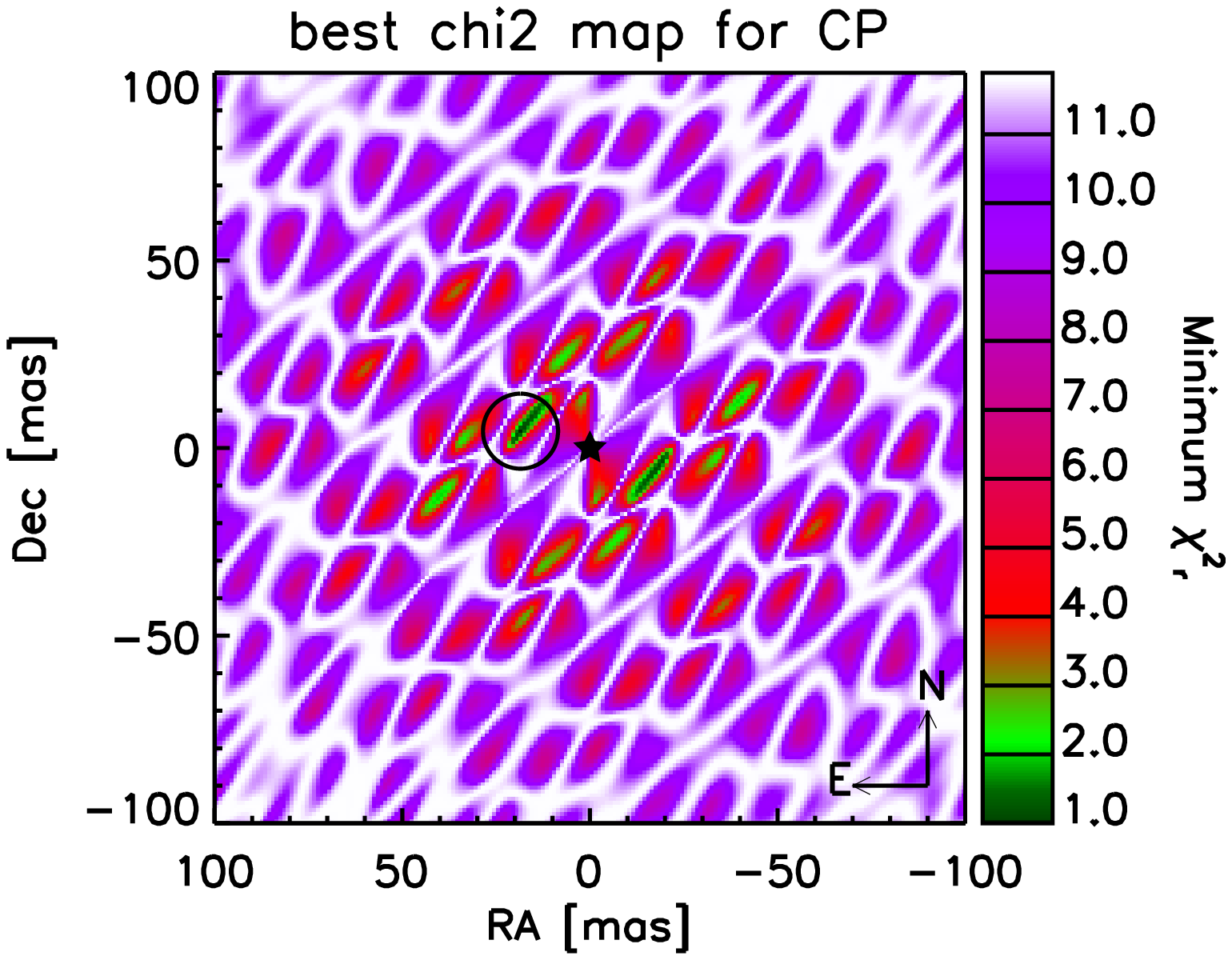}
\includegraphics[scale=0.33]{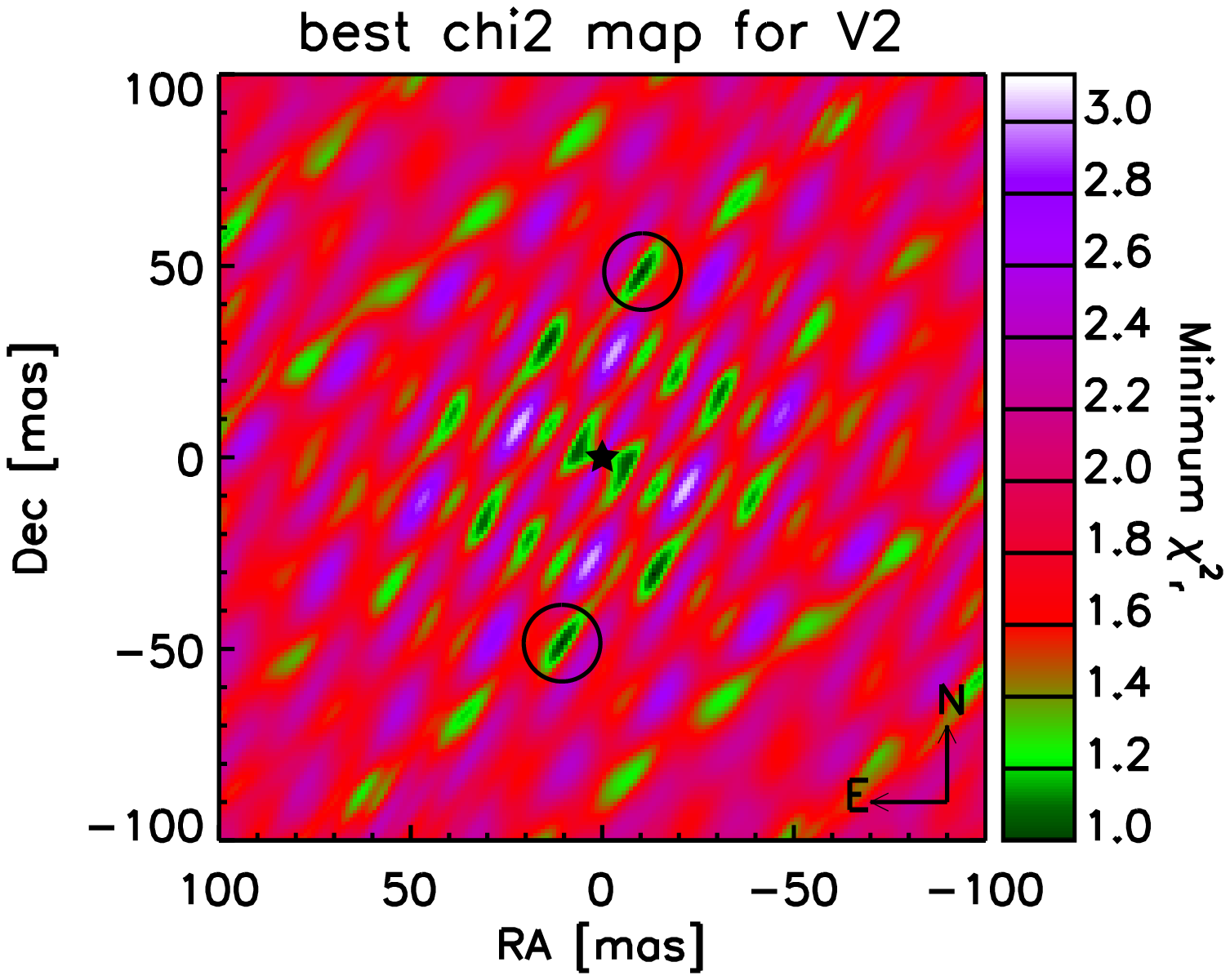}
\caption{Normalised $\chi^2$ maps of the combined CP+$V^2$ (left), the CP alone (middle) and the $V^2$ alone (right), for the five stars identified as having an off-axis companion: HD~4150, HD~16555, HD~29388, HD~202730, and HD~224392 (from top to bottom). The black circles indicate the positions of the minima in the maps. Two identical minima always appear on both sides of the star for the $V^2$, because of the well-known $180\degr$ ambiguity.}
 \label{fig:chi2maps}
\end{figure*}

To determine whether or not a companion is detected, we use, as explained in Sect.~\ref{sub:principle}, the significance level of the detection, which can be associated with a given confidence level if the underlying probability distribution function is known. The threshold above which a detection will be identified must however be chosen with care. A 3$\sigma$ significance level was used by \citet{Absil11}, following common practice in presence of Gaussian noise. This choice is however not straightforward, and can be backed up by studying the noise properties of the data set. This can conveniently be done by including negative contrasts for the off-axis companions in the definition of our combined $\chi^2$ cube. While non physical, negative companions can be used to represent positive fluctuations in the $V^2$ (i.e.\ situations where the measured $V^2$ is higher than the expected $V^2$ from the photosphere). Negative companions do not significantly affect the CP, which can take both positive and negative values. A negative companion would indeed produce the same CP signature as a positive companion located on the opposite side of the star. In the following, we associate negative significance levels with negative companions. The histogram of the significance levels for our complete sample (listed in Table~\ref{tab:all}) is illustrated in Fig.~\ref{fig:histosignif}, where the range of the plot has been limited to $[-10,10]$ for the sake of clarity. The negative part of the histogram does not contain any astrophysical signal, and can therefore be used as a reference to study the noise properties of our sample. The absence of significance levels close to 0$\sigma$ in the histogram can be explained by the fact that, in presence of noise and because of the limited number of observations, it is always possible to obtain a better fit to our data sets by inserting a companion somewhere in the field of view rather than using the single-star model (a significance level of 0 would only happen if the data were in perfect agreement with the single-star model).

\begin{table*}[t]
\caption{Main properties of the newly detected companions and their host stars (excluding HD~224392, which will be discussed in a forthcoming paper).}
\label{tab:companions}
\centering
\begin{tabular}{ccccccccccc}
\hline \hline
 & & & & & \multicolumn{2}{c}{Star} & & \multicolumn{3}{c}{Companion} \\ \cline{6-7} \cline{9-11}
Name  & R.A. & Dec. & Distance & & Spectral  & $H$ & & Separation & P.A. & Contrast \\
& & & (pc)  & & type & (mag) & & (mas) & (deg) & (\%) \\
\hline
HD~4150 & 00:43:21 & $-$57:27:47 & 75.5 & & A0IV & $4.35\pm0.08$ & & $90.5 \pm 2.2$ & $84.0\pm 2.2$ & $2.3 \pm 0.4$\\
HD~16555  & 02:37:24 & $-$52:32:35 & 45.6 & & A6V & $4.60\pm0.02$ & & $78.7 \pm 1.6$ & $40.9\pm 0.3$ & $51 \pm 4$ \\
HD~29388  & 04:38:09 & $+$12:30:39 & 47.1 & & A6V & $3.94\pm0.01$ & & $11.1 \pm 0.2$ & $71.6 \pm 0.05$ & $3.1 \pm 0.2$ \\
HD~202730 & 21:19:51 & $-$53:26:57 & 30.3 & & A5V & $4.22\pm0.07$ & & $61.7 \pm 1.2$ & $-21.4 \pm 0.8$ & $87 \pm 14$ \\
\hline
\end{tabular}
\tablefoot{The H magnitudes are taken from 2MASS \citep{Skrutskie06}, except in the case of HD~29388, for which the magnitude is from \citet{Kidger03}.}
\end{table*}

Out of the 92 stars in our sample, 38 show a negative significance level, while 54 show a positive significance level (see Fig.~\ref{fig:histosignif} and Table~\ref{tab:all}). This suggests that around 16 stars in our sample have circumstellar emission emanating from a disc or a companion. This is consistent with the nine hot exozodiacal discs identified by \citet{Ertel14} and the six companions described in the next section. From the 38 stars with a negative significance level, only three are located below $-3 \sigma$. While the distribution of the significance level looks far from Gaussian in Fig.~\ref{fig:histosignif}, we decide to use $3 \sigma$ as our detection criterion. From the negative part of the histogram, we expect that three of the thirteen stars showing a significance level higher than $3 \sigma$ could be false positive detections. All thirteen stars will be carefully inspected to assess the possible presence of such false positives.

	\subsection{Results of the search} \label{sub:results}

Table \ref{tab:detections} lists all the stars that have a significance level higher than $3\sigma$ for the combined $\chi^2$ analysis. As already mentioned in Sect.~\ref{sub:principle}, by looking separately at the $\chi^2$ maps for the CP and $V^2$, we can readily discriminate between a companion and a disc. This is done in Table~\ref{tab:detections}, where the significance levels are given based on the analysis of the CP and of the $V^2$ separately. The eight stars having a significance level below $3\sigma$ in the analysis of the CP, while showing a significant detection in the $V^2$ analysis, are understood to be surrounded by a circumstellar disc rather than a companion. Seven of these stars are indeed among nine stars identified by \citet{Ertel14} as having a bright exozodiacal disc. A special note must be added to HD~15798, which was very recently shown by \citet{Tokovinin14} to host a faint companion at an angular distance of $0\farcs21$, thanks to speckle interferometry observations. This companion, located outside the PIONIER search region, could not have been detected with the search method used here, as the fringe packets of the two stars are separated on most baselines (or even not within the OPD scanning range on some baselines). Its contribution is therefore considered mostly as incoherent light in the PIONIER data reduction pipeline. With an I-band contrast of four magnitudes, this companion is most probably at the origin of the $V^2$ drop observed by PIONIER for HD~15798. It is difficult to derive the H-band contrast of this companion from our observations, because of the way its contribution to the coherent flux is handled by the PIONIER pipeline, and because it is significantly affected by the off-axis transmission profile of the PIONIER instrument.

The five remaining stars (HD~4150, HD~16555, HD~29388, HD~202730 and HD~224392) are identified as having an off-axis companion. This classification is backed up by a careful inspection of the $\chi^2$ maps separately for the CP and $V^2$: to validate the detection of an off-axis companion, the two maps should have their minima (or sufficiently deep local minima) at the same position. This is the case for four of our five newly identified binary stars (see Fig.~\ref{fig:chi2maps}). Only HD~224392 shows a somewhat suspicious behaviour, where the various $\chi^2$ maps do not show local minima at the same position (see bottom row of Fig.~\ref{fig:chi2maps}). This star will be the subject of a forthcoming study (S.~Borgniet et al., in prep), where the binarity of the source is confirmed by further observations. We will therefore not make further comments on this star in the present manuscript. Two additional stars show a potential signature of companion: HD~20794 and HD~23249. They both have a significance higher than $3\sigma$ in their combined $\chi^2$ map for one of two epochs. A closer inspection of the individual $\chi^2$ maps reveals that the positions of the minima do not match, which indicate that the excess emission is most probably due to a circumstellar disc rather than a point-like companion.

\begin{figure*}[t]
\centering
\includegraphics[scale=0.33]{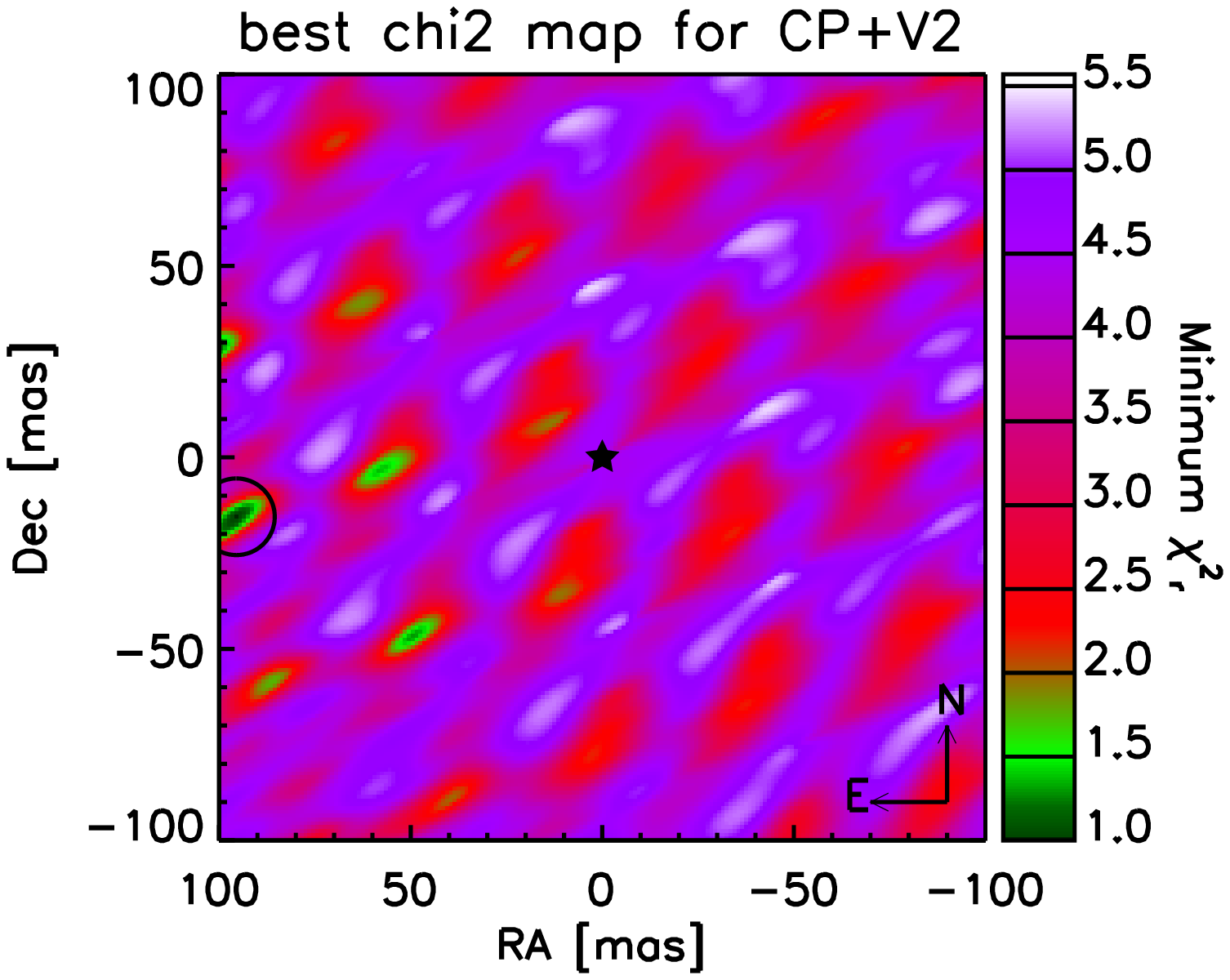}
\includegraphics[scale=0.33]{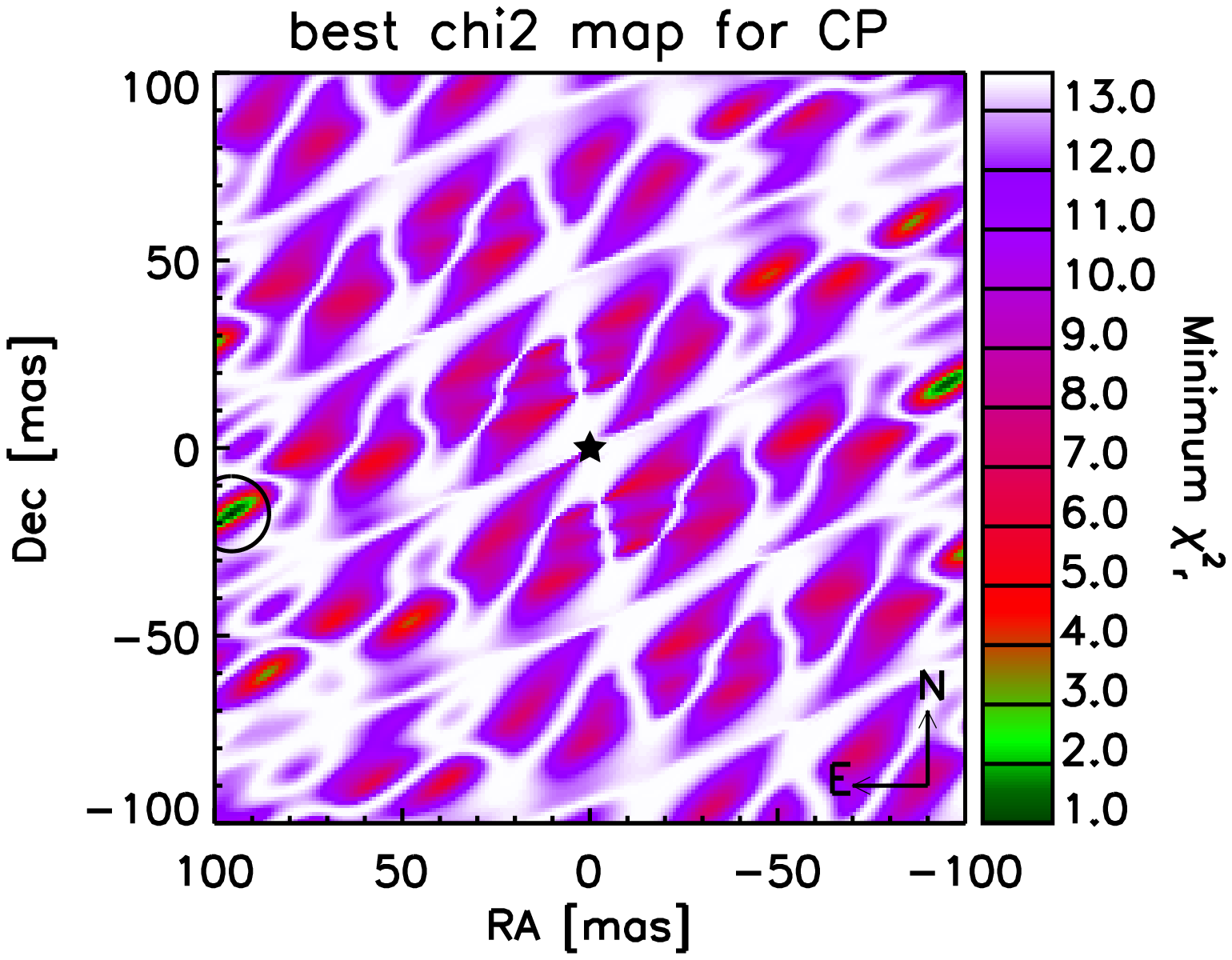}
\includegraphics[scale=0.33]{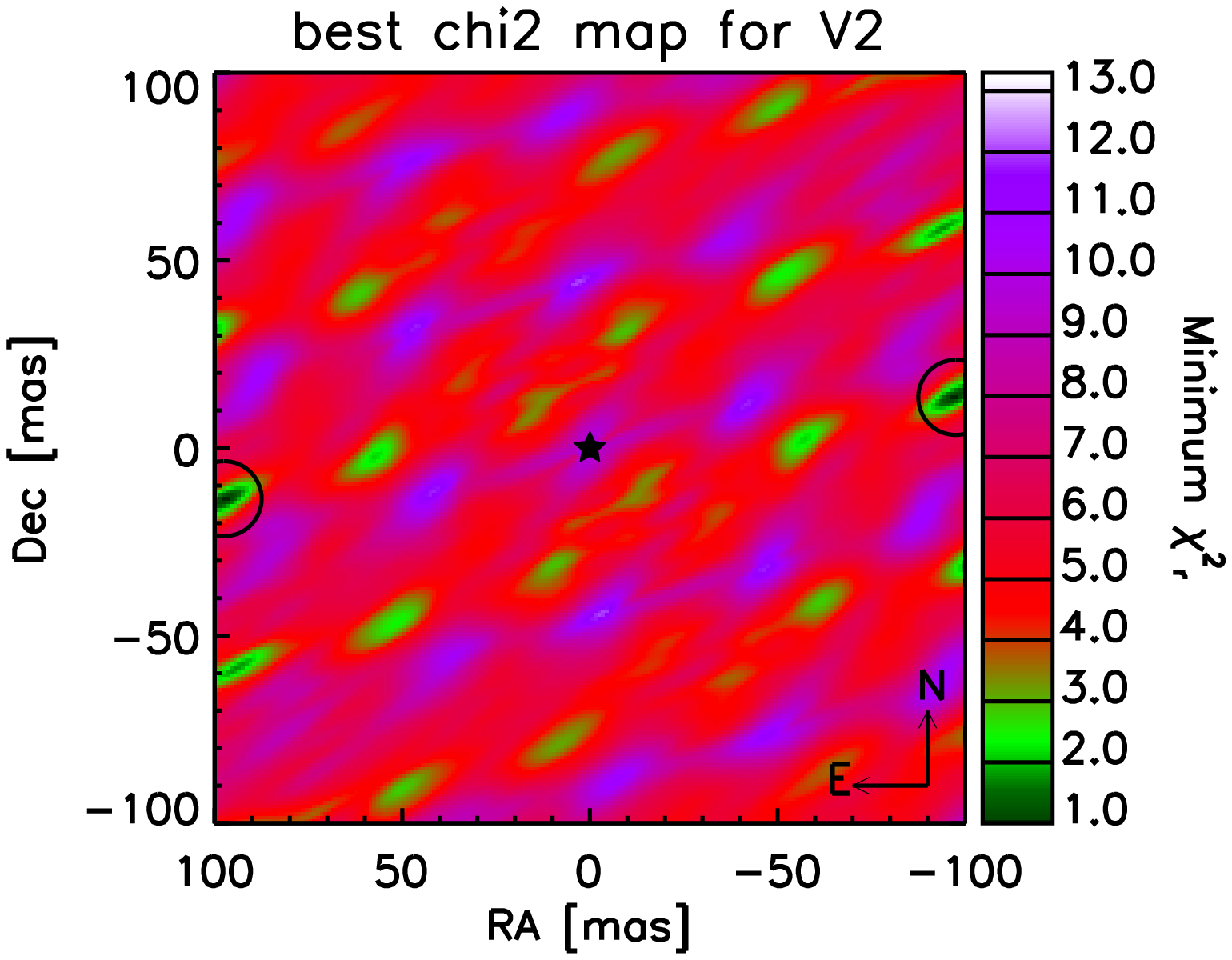}
\caption{Same as Fig.~\ref{fig:chi2maps} for the second epoch observation of HD~4150, obtained on 9 August 2013 with PIONIER operated in K band on the compact AT configuration.}
 \label{fig:HD4150}
\end{figure*}

A summary of all the newly found companions (excluding HD~224392) can be found in Table~\ref{tab:companions}, where we also give the contrast and position of the detected companions. Surprisingly, two of them show a best-fit contrast of 0.5 or larger. The reason why such bright companions had not been revealed to date by other observations will be discussed in Sects.~\ref{sec:properties} and \ref{sec:discussion}.

%__________________________________________________________________

\section{Properties of the detected companions}
\label{sec:properties}

Here, we describe the properties of the newly found companions and of their host stars, and we give a rough estimation of the orbital period. It is indeed most probable that the detected companions are gravitationally bound to their host star. The probability of finding such bright background (or foreground) objects inside the tiny PIONIER field of view can be computed by evaluating the density of stars of magnitude inferior or equal to the detected companions in the immediate neighbourhood of the primary stars. We estimate this density by querying the 2MASS catalogue \citep{Skrutskie06} within a radius of $1\degr$ around the four stars: the probability of finding an unrelated background object as bright as the detected companions within the PIONIER field of view is never larger than a few $10^{-8}$. 

	\subsection{HD~4150}

Also known as eta Phe, HD~4150 (HIP~3405, HR~191) is an A0IV star surrounded by a cold debris disc \citep{Su06}. It belongs to a visual double system according to the WDS \citep{Mason01} and the CCDM \citep{Dommanget02} catalogues. The previously detected visual companion is faint \citep[$V=11.5$,][]{Dommanget00} and located at about $20\arcsec$ from the primary. It is therefore not expected to affect our interferometric measurements in any way. In addition, HD~4150 is classified as an astrometric binary based on the comparison of the Tycho-2 and Hipparcos astrometric catalogues \citep{Makarov05,Frankowski07}, which show discrepant proper motions. The star also shows an acceleration of $-6.41\pm0.99$~mas/yr$^2$ in right ascension and $1.59 \pm 0.98$~mas/yr$^2$ in declination as measured by Hipparcos \citep{Makarov05}. No orbital solution is given, however, and the nature of the companion is unknown.

Two important properties can be estimated from the contrast and angular separation found in Sect.~\ref{sub:results}: the spectral type and the orbital period. We derive the spectral type by assuming a main sequence companion of lower mass than the primary. This is most likely, as a more massive, post-main sequence companion would be brighter than the primary, while a white dwarf would be much fainter and would remain undetected. The H-band flux ratio amounts to $0.023 \pm 0.004$, which corresponds to $\Delta H = 4.1 \pm 0.2$. Taking the magnitude and distance of HD~4150 into account, the companion has an absolute magnitude $M_H=4.1 \pm 0.2$, which corresponds roughly to a K1V star according to \citet{Allen00}. A lower bound can then be given on the period of the orbit, assuming a face-on system. The measured angular separation corresponds to a linear separation of 6.8~AU. Using Kepler's third law and an estimated mass of 2.82 $M_{\odot}$ for the primary \citep{Gerbaldi99}, we derive a minimum period of about 9.5~years. This estimated period is consistent with an astrometric detection of the companion \citep{Makarov05}. It is therefore likely that the companion detected within the present study is the source of the astrometric signature.

To monitor the orbital movement of the companion, a second observation of the system was obtained on 9 August 2013 with PIONIER operated at K band on the same AT configuration (programme ID 091.C-0597). The companion is recovered with a contrast of $0.041\pm0.004$ at an angular separation of $96.8\pm2.4$~mas and position angle of $-99.2\degr \pm 1.1\degr$ (see Fig.~\ref{fig:HD4150}). The increase in the angular separation reveals that the orbit is not face on. Nevertheless, assuming a face-on orbit and a minimum period of 9.5~years, we estimate that the companion should have moved by about $24\degr$ on its orbit. This is of the same order as the difference in position angle between the two epoch ($15\fdg2$). We conclude that the companion is most probably on a slightly inclined and/or eccentric orbit with a period slightly longer than 10~years. We note that the measured K-band contrast points towards a slightly more massive companion, with a spectral type around G5V. More observations will need to be performed to nail down the orbital parameters and spectral type of the companion.

	\subsection{HD~16555}

Eta Hor (HD~16555, HIP~12225, HR~778) is an A6V star classified as a stochastic astrometric binary by \citet{Frankowski07}. Despite the stochastic character of the astrometry (i.e.\ the impossibility of finding a suitable orbital solution based on the Hipparcos astrometry), \citet{Goldin07} propose an orbital solution with a semi-major axis of $23.1^{+1.6}_{-1.0}$~mas and period of 3.0~yr. No information can however be derived on the nature of the companion. Later on, eta Hor was observed with adaptive optics, showing only a possible (yet unconfirmed) faint companion at $4\farcs9$ \citep{Ehrenreich10}, which is not expected to affect our interferometric observations in any way. More recently, the star was resolved as a binary by speckle interferometry \citep{Hartkopf12}, showing the presence of a bright companion ($\Delta I=0.7$~mag, $\Delta y=1.4$~mag) at an angular separation of 70.3~mas and a position angle of $60\fdg4$ (Besselian epoch of observation: 2011.0393~yr). This publication came out after we finalised the sample selection for the \textsc{Exozodi} survey, which explains why this star was nonetheless observed during our survey.

The companion found in our observations, showing a contrast $\Delta H \simeq 0.7$ at an angular separation of 78.7~mas (i.e.\ projected linear separation of 3.6~AU) and position angle of $40\fdg9$, seems to match well the \citet{Hartkopf12} discovery. It however hints at a significantly eccentric and/or inclined orbit. Although the current astrometric information is too scarce to confirm it, this visual binary could also match the stochastic astrometric solution proposed by \citet{Goldin07}. A face-on, circular orbit with a semi-major axis of 3.6~AU would indeed have a period of about 3.7~yr, which is in line with the 3.0~yr period found by \citet{Goldin07}.

Finally, we can infer the spectral type of the companion from the measured contrast. With an absolute H-band absolute magnitude $M_H=1.31 \pm 0.02$ for the primary and a contrast $\Delta H = 0.73\pm 0.08$, the companion has an estimated absolute magnitude $M_H=2.04 \pm 0.08$, which gives an F0V spectral type according to \citet{Allen00}.

	\subsection{HD~29388}

90 Tau (HD~29388, HIP~21589, HR~1473) is an A6V member of the Hyades cluster. It hosts a cold debris disc \citep{Su06}. Although it was observed by speckle interferometry \citep{Patience98}, no sign of binarity was found so far around that star to our knowledge, despite being extensively observed since the 1950s as a chemically peculiar member of an open cluster. 90 Tau is nevertheless listed as a member of a double system in Simbad, but with only a faint visual companion at a distance of about $120\arcsec$. The companion discovered here is thus completely new.

With an absolute magnitude $M_H=0.71\pm0.01$ for its primary star and a contrast $\Delta H=3.77\pm0.07$, the companion has an absolute magnitude $M_H=4.48\pm0.07$, which corresponds to a K4V star according to \citet{Allen00}. Assuming a face-on, circular orbit with a semi-major axis of 11.1~mas (0.52~AU), the period would be around 84~days. This can be considered as a minimal period for the system.

	\subsection{HD~202730}

Tet Ind (HD~202730, HIP~105319, HR~8140) has been known to host a visual G0V companion (GJ~9733~B) since the 19th century. Its angular separation has been measured over 150~years and is well approximated with a linear motion in the Catalog of Rectilinear Elements\footnote{http://www.usno.navy.mil/USNO/astrometry/optical-IR-prod/wds/lin1}. In July 2012, the companion is expected to be separated by about $7\farcs2$ from the primary, which is comfortably outside the PIONIER field of view (and even outside the AT field of view). Tet Ind has also been shown to have a constant radial velocity in the survey of \citet{Lagrange09}. The discovery of a nearly equal flux companion therefore comes as a surprise.

Based on the measured contrast ($\Delta H = 0.15 \pm 0.17$), we estimate the companion to have an A5V spectral type like the primary. The odds for a face-on orbit are high (no radial velocity variations detected). Assuming a circular orbit, the period would then be around 1.3~years.

%__________________________________________________________________

\section{Discussion} \label{sec:discussion}

	\subsection{A population of undetected stellar companions?}

Five binary stars were identified in this study. In addition to the four stars discussed in Sect.~\ref{sec:properties}, we include here HD~224392, whose companion will be discussed in a forthcoming paper (S.~Borgniet et al., in prep.). This is an A-type star, just as the four others are. The fact that five stars among the 30 supposedly single A-type stars\footnote{although we recognise that two of them have been classified as astrometric binaries in the literature} of the \textsc{Exozodi} sample were revealed to be binaries deserves some further discussion.

\begin{figure*}[t]
\begin{center}
\includegraphics[scale=0.33]{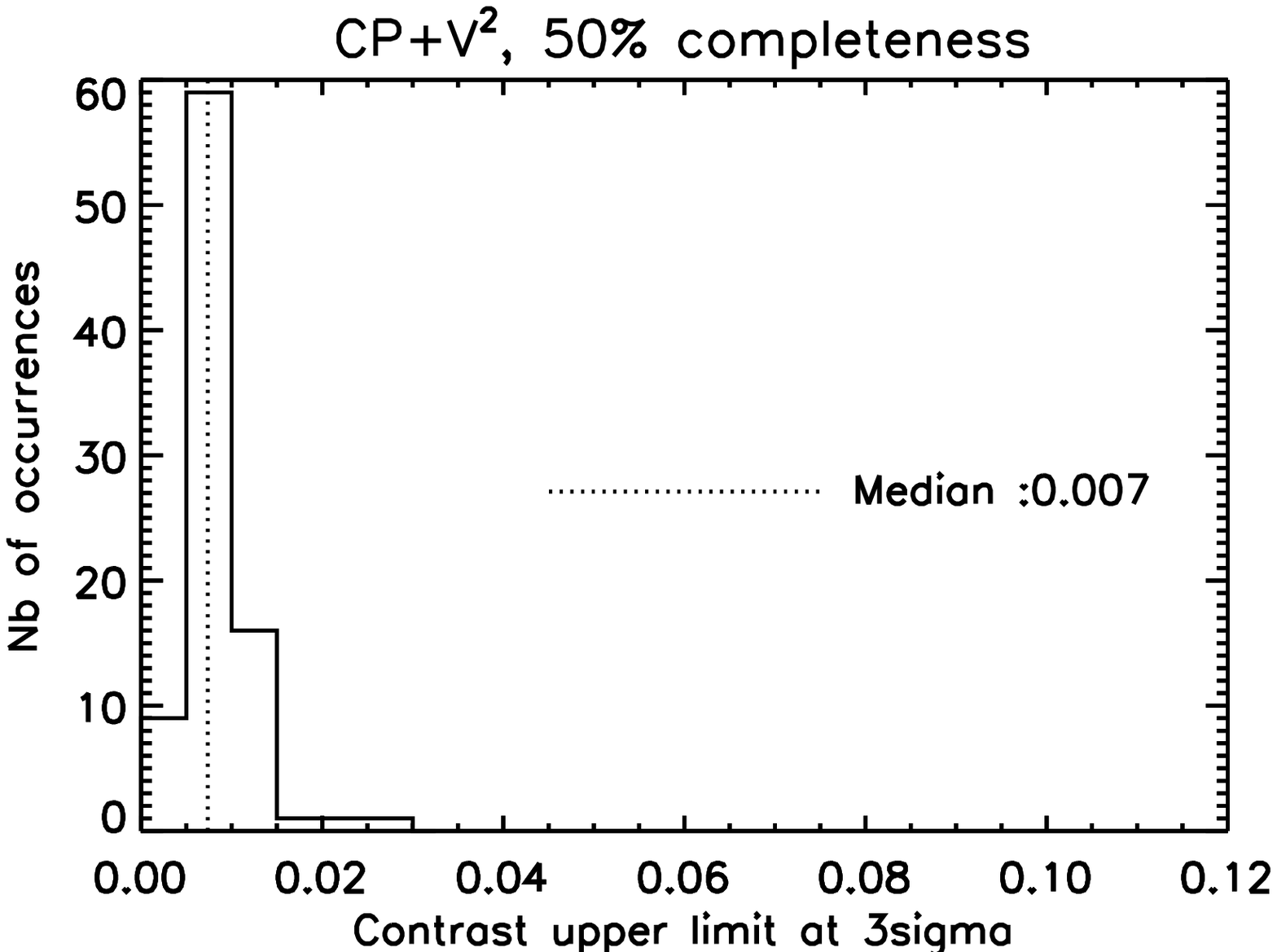}\quad
\includegraphics[scale=0.33]{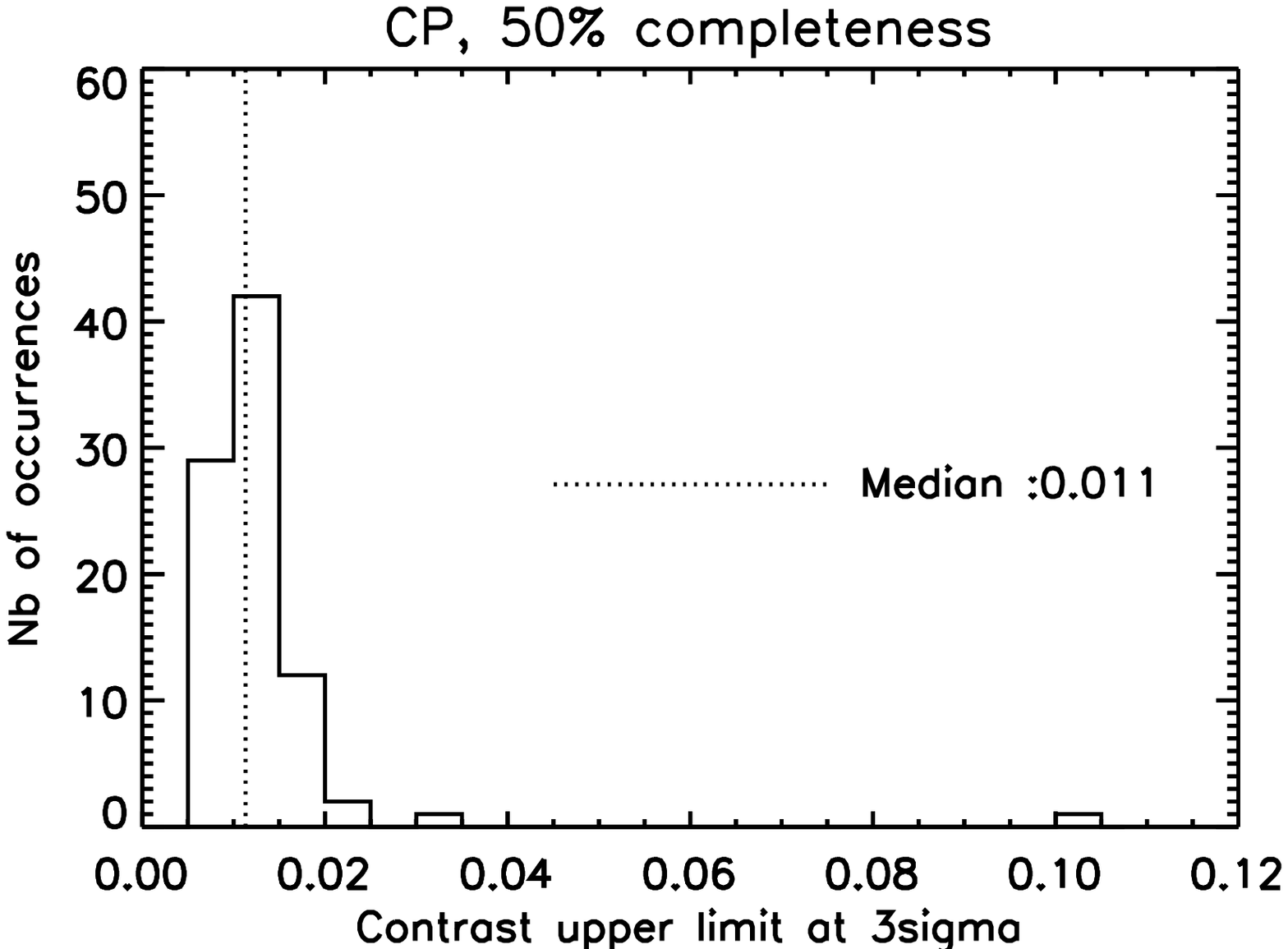}\quad
\includegraphics[scale=0.33]{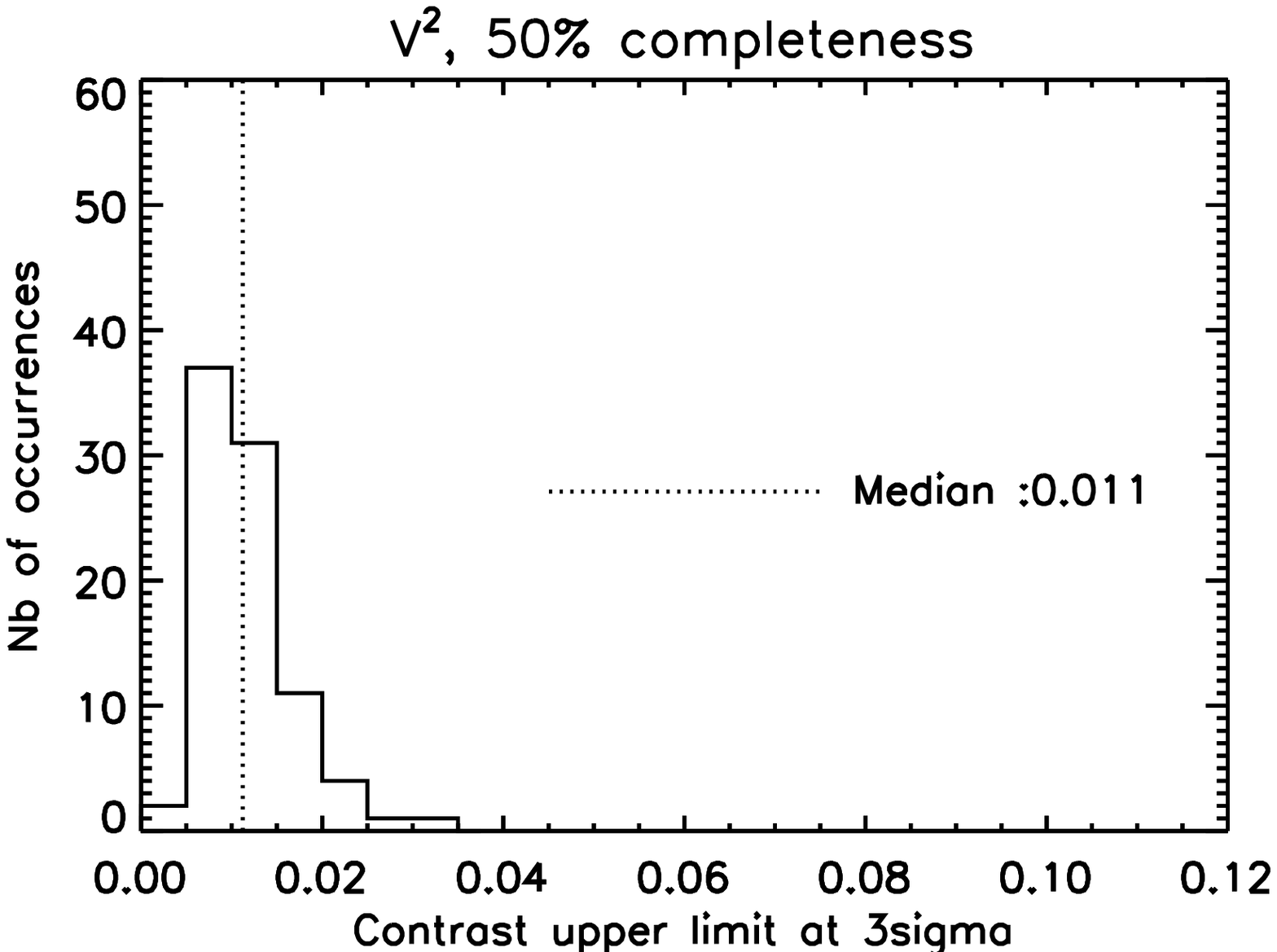}\\ \vspace*{1mm}
\includegraphics[scale=0.33]{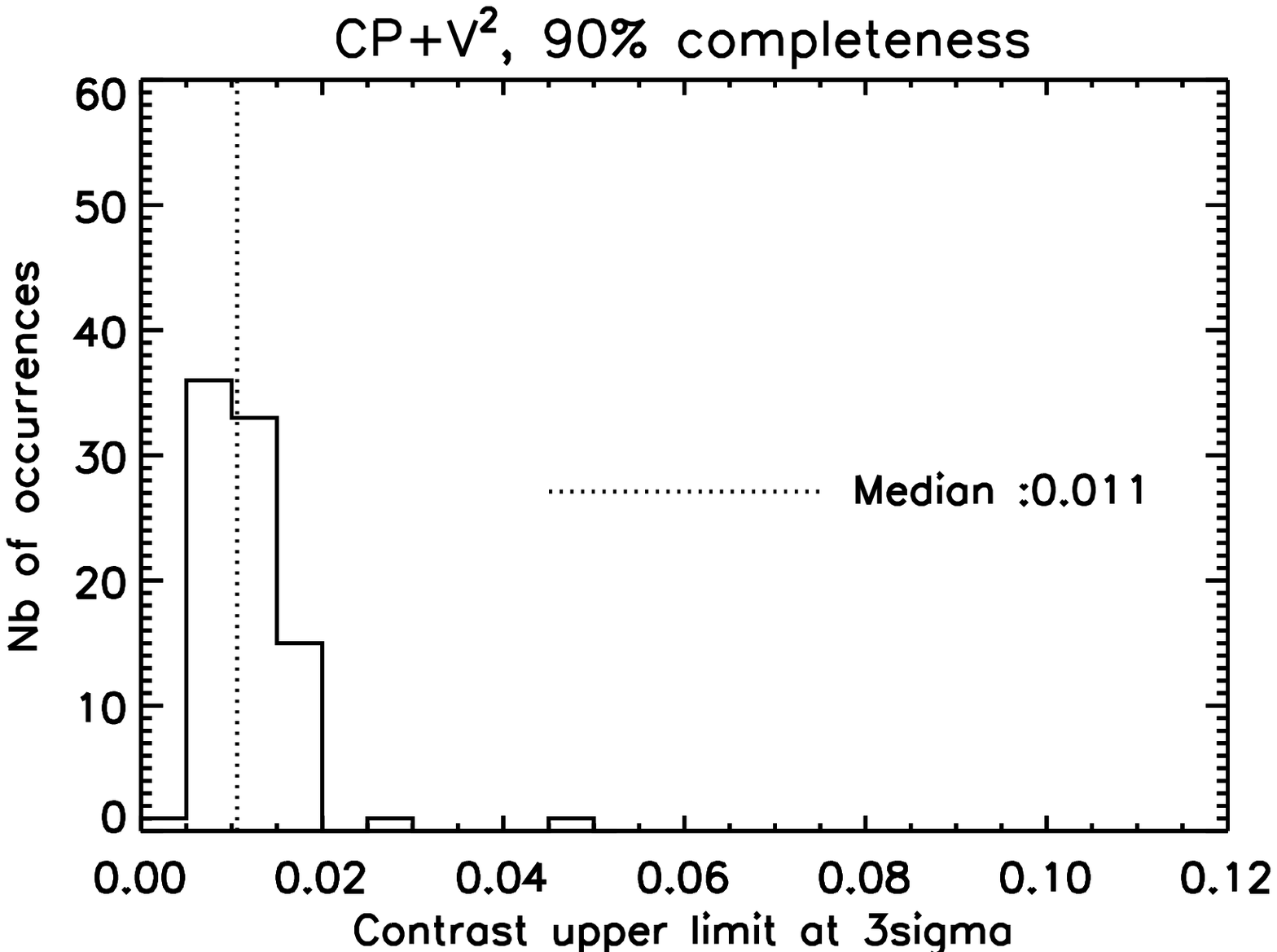}\quad
\includegraphics[scale=0.33]{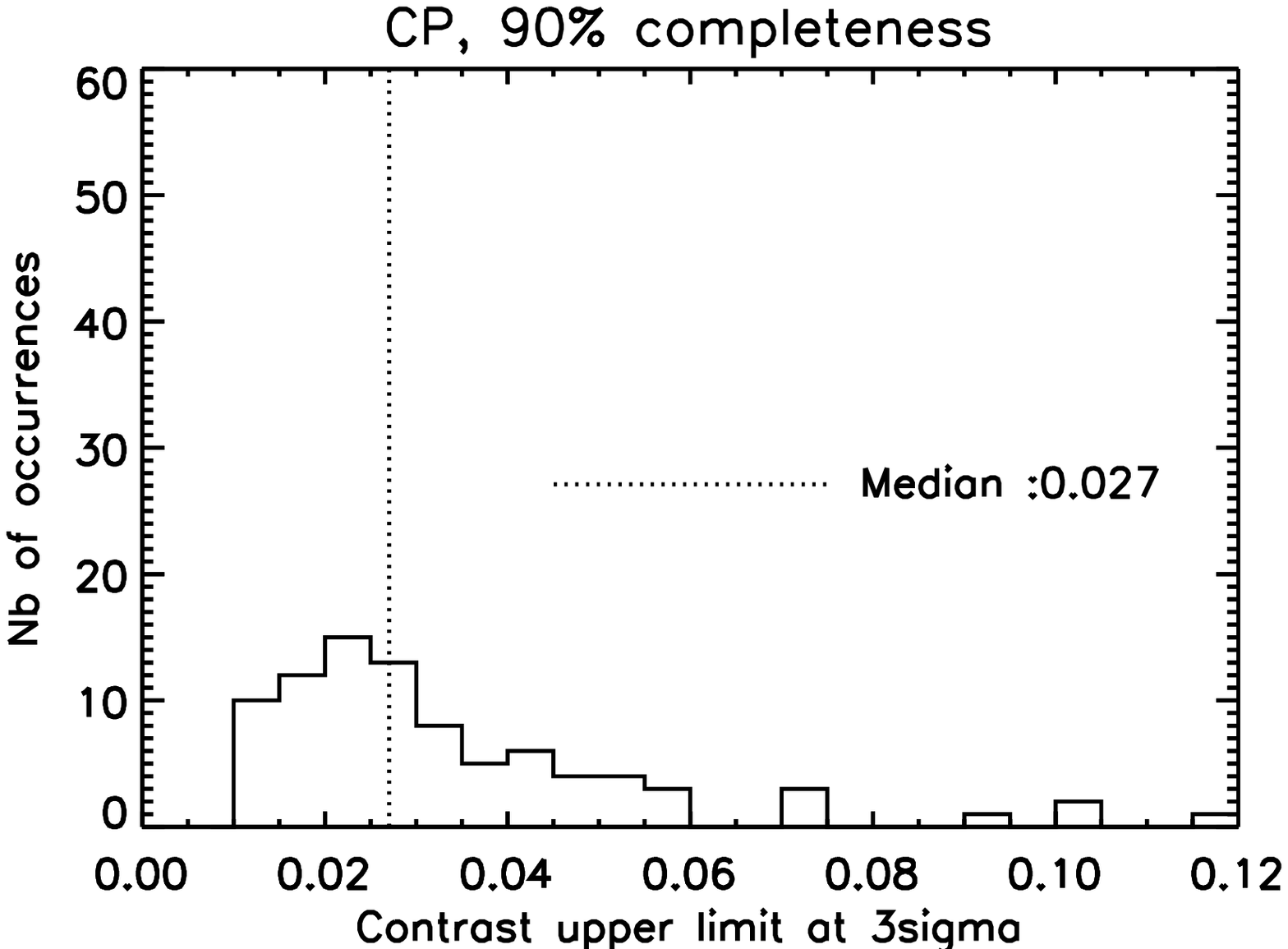}\quad
\includegraphics[scale=0.33]{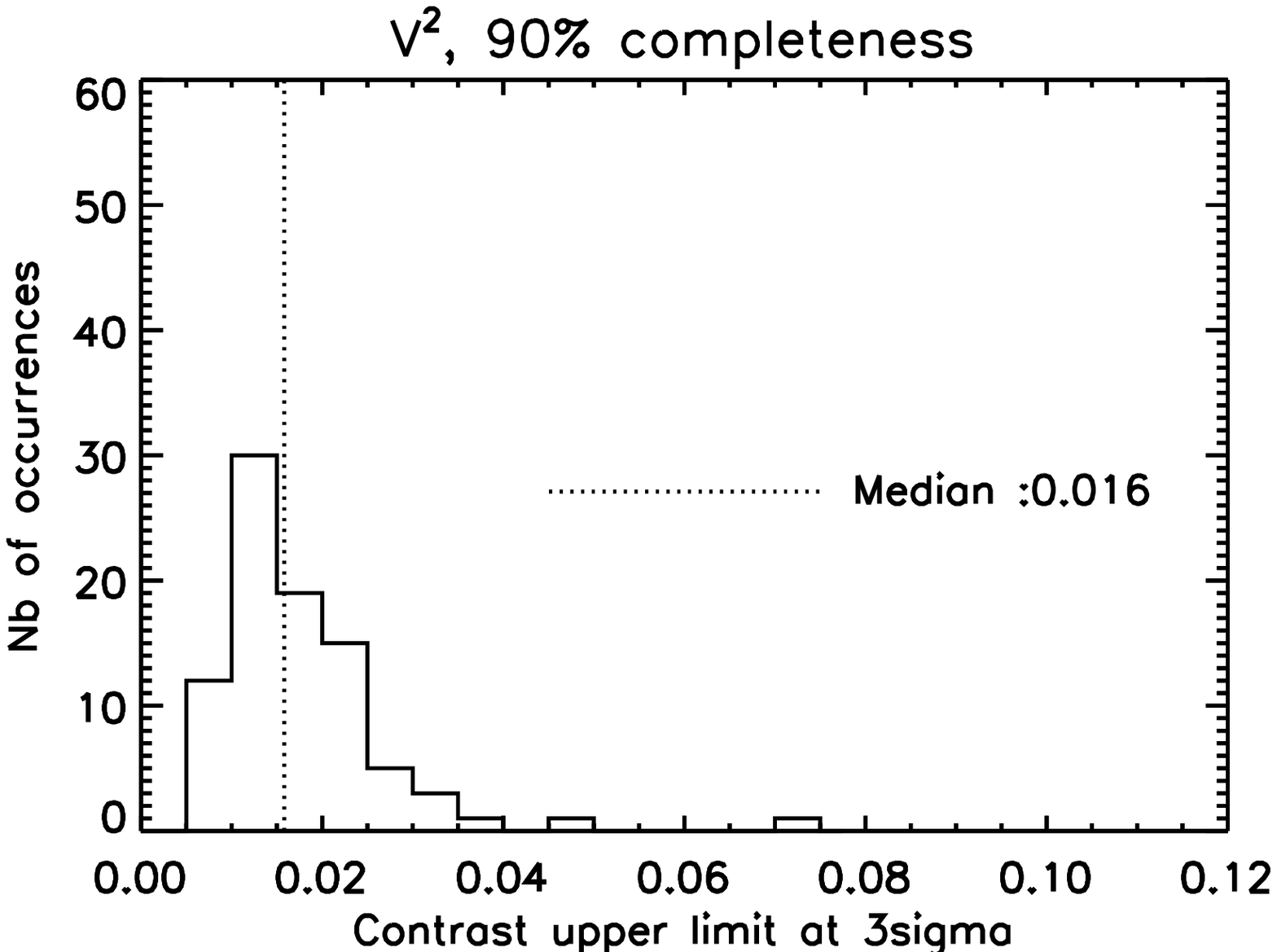}\\
\end{center}
\caption{Histogram of the $3\sigma$ contrast upper limit for 50\% (top) and for 90\% (bottom) of the field of view, for the  combination of CP and $V^2$ (left), the CP only (middle), and the $V^2$ only (right).}
 \label{fig:sensitivitydet}
\end{figure*}

Historically, many dedicated studies have been carried out to quantify the number of binaries in the solar neighbourhood, showing higher multiplicity rates around more massive stars \citep[see][for a recent review]{Duchene13}. A lot of papers were published regarding the multiplicity of solar-type stars (F-G stars), including e.g.\ censuses of binaries within 25~pc \citep{Raghavan10} and within 67~pc \citep{Tokovinin14I,Tokovinin14II}. Later spectral types have also been extensively addressed, including a quasi-complete sample within 10~pc in \citet{Henry06}. Even brown dwarfs have had their share of multiplicity studies \citep[see][for a review]{Burgasser07}. However, when it comes to earlier spectral types, and A-type stars in particular, assessing the multiplicity is much more challenging. One of the main reasons is that classical spectroscopic methods do not work very well in the case of A-type stars, for which the spectral lines are generally broadened by high rotational velocities (typically 100--200~km~s$^{-1}$) and usually blended. Spectroscopic surveys of A-type stars are therefore much less sensitive and complete than for solar-type stars. They have mostly targeted chemically peculiar A-type stars so far \citep[e.g.][]{Carrier02,Carquillat07}. A new method for radial velocity measurements has only recently been proposed to survey intermediate-mass stars \citep{Galland05}, enabling the search for substellar companions, although with an RV accuracy much worse than for solar-type stars. 

Systematic searches for visual companions around A-type stars also face a significant challenge, i.e.\ the steep mass-luminosity relationship, which makes low-mass companions hard to detect. Only recently have large-scale, dedicated (AO-assisted) surveys been conducted, including most notably the VAST survey \citep{DeRosa11}. The first results of this survey, targeting an incomplete sample of 435 A-type stars out to 75~pc, show that $33.8\pm2.6\%$ of A-type stars are visual binaries in the separation range between 30 to 10\,000~AU (i.e.\ about $0\farcs4$ to $140\arcsec$). A significant gap however remains between the realm of spectroscopic binaries and the visual binary domain. The study of astrometric binaries from the Hipparcos catalogue partly fills that gap, but the nature of the detected companions, and even their orbital parameters, are generally not well constrained. Interferometry offers an attractive way to fill that gap, as recently proposed for massive stars \citep{Sana14}, before GAIA revolutionises the field.

With the present study, we address a separation range (1--100~mas) that has only been partially scraped by speckle interferometry in the past \citep[e.g.][]{Mason13,Tokovinin14}. Among the five stars resolved as binaries within the present survey, only one was unambiguously identified as a visual, gravitationally bound, binary before the start of our survey (HD~202730, which has a companion at $7\farcs2$). The four others are new visual companions, although the companion to HD~16555 was independently published by \citet{Hartkopf12}. We can therefore consider that four out of 30 (i.e.\ $13.3_{-4.0}^{+8.6}\%$) A-type stars in our sample revealed to be unknown visual binaries. Adding this to the previous result of \citet{DeRosa14} suggest that about 47\% of A-type stars are visual binaries for angular separations ranging from 1~mas to $140\arcsec$, although a gap remains between the outer edge of the PIONIER field of view ($\sim 100$~mas) and the inner edge of the \citet{DeRosa14} survey ($0\farcs4$). This gap could adequately be covered by more systematic speckle interferometry or sparse aperture masking observations. For instance, as explained in Sect.~\ref{sub:results}, the F5V star HD~15798 was recently shown to host a faint companion at an angular distance of $0\farcs21$ by \citet{Tokovinin14}.

Based on this discussion, it is legitimate to ask oneself how many more close companions remain undiscovered around A-type stars in the solar neighbourhood. Our study suggests that, even after a careful target selection, five out of 30 (i.e.\ 17\%) A-type stars still have a stellar companion within an angular separation range that critically affects interferometric observations. Using A-type stars as calibrators for interferometric observations should therefore be avoided as much as possible. We also note that, despite our efforts to search the literature for the signs of faint companions, some of the hot exozodiacal disc detections claimed within our CHARA/FLUOR survey of debris disc stars \citep{Absil13} could also be due to unknown, close companions. Assuming that up to 17\% of the observed A-type stars in the CHARA/FLUOR survey have a unknown faint companion within the FLUOR field of view (similar to the PIONIER fleld of view) would still preserve a hot exozodiacal disc occurrence rate significantly larger around A-type stars ($\sim$33\%) than solar-type stars ($\sim$18\%) in that survey.

	\subsection{On the PIONIER sensitivity}
	\label{sub:pionier}

In the case of non detections (79 stars out of 92 in our sample), we can determine an upper limit on the presence of a companion as a function of the position in the field of view, using the $\chi^2$ statistics as explained in \citet{Absil11}. The resulting sensitivity map can then be used to derive the median sensitivity (i.e.\ sensitivity achieved for 50\% of the positions within the search region). We can also define the sensitivity at a higher completeness level, e.g.\ the sensitivity reached for 90\% of the positions within the search region. In order to deduce the typical sensitivity of PIONIER in survey mode  (3 OBs per target), we produced the histograms of the sensitivity levels for all stars showing no near-infrared excess. These histograms give us the sensitivity to companions in 50\% or 90\% of the search region at a significance level of $3\sigma$ for the CP, the $V^2$, and the combination of the two. They are illustrated in Fig.~\ref{fig:sensitivitydet}. We deduce from this figure that the sensitivity of PIONIER is around 1\% when using the CP and the $V^2$ in a combined way (median sensitivity of 0.7\%, percentile 90 sensitivity of 1.1\%).

\begin{figure*}[t]
\centering
 \includegraphics[scale=0.33]{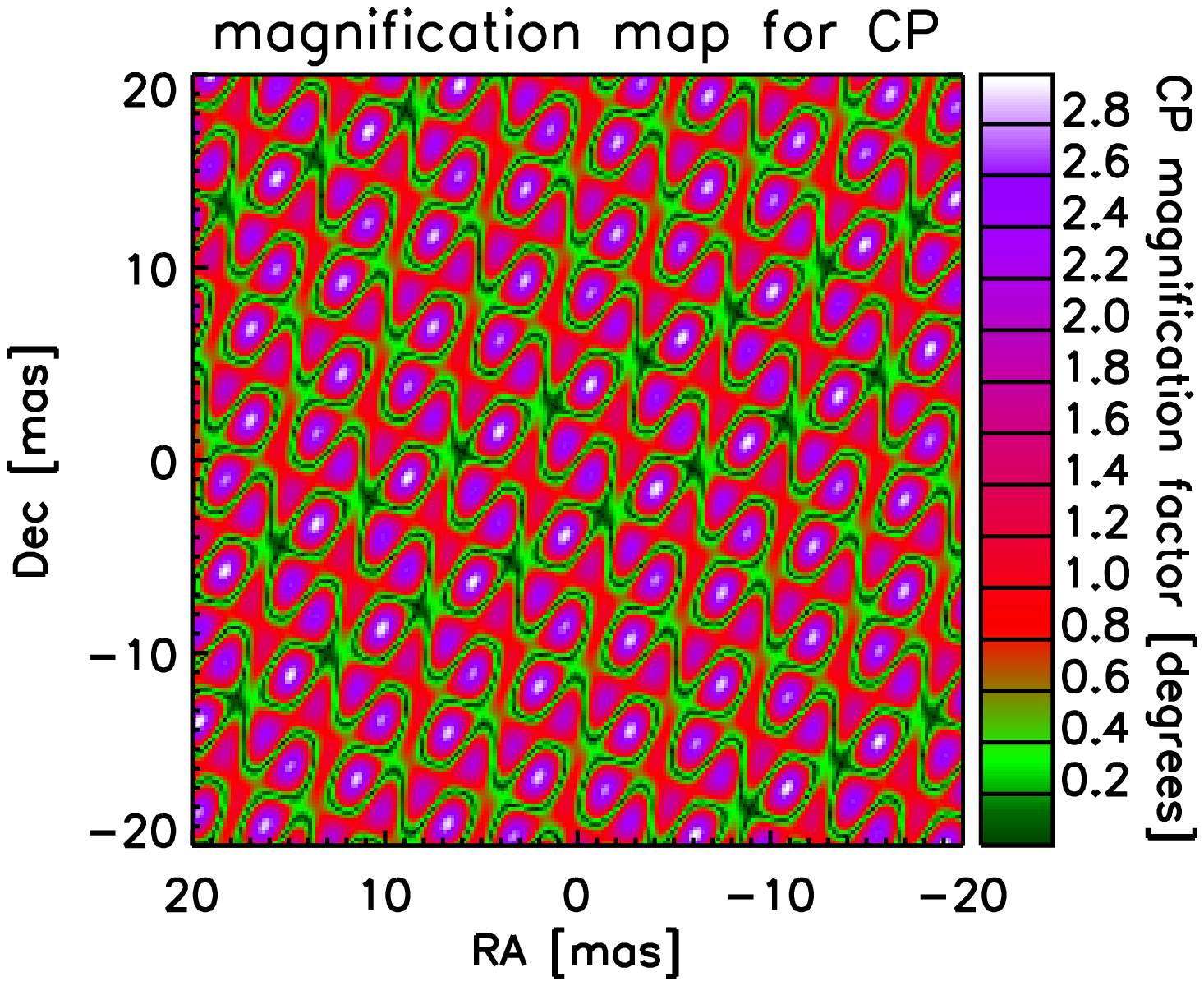}\quad
\includegraphics[scale=0.33]{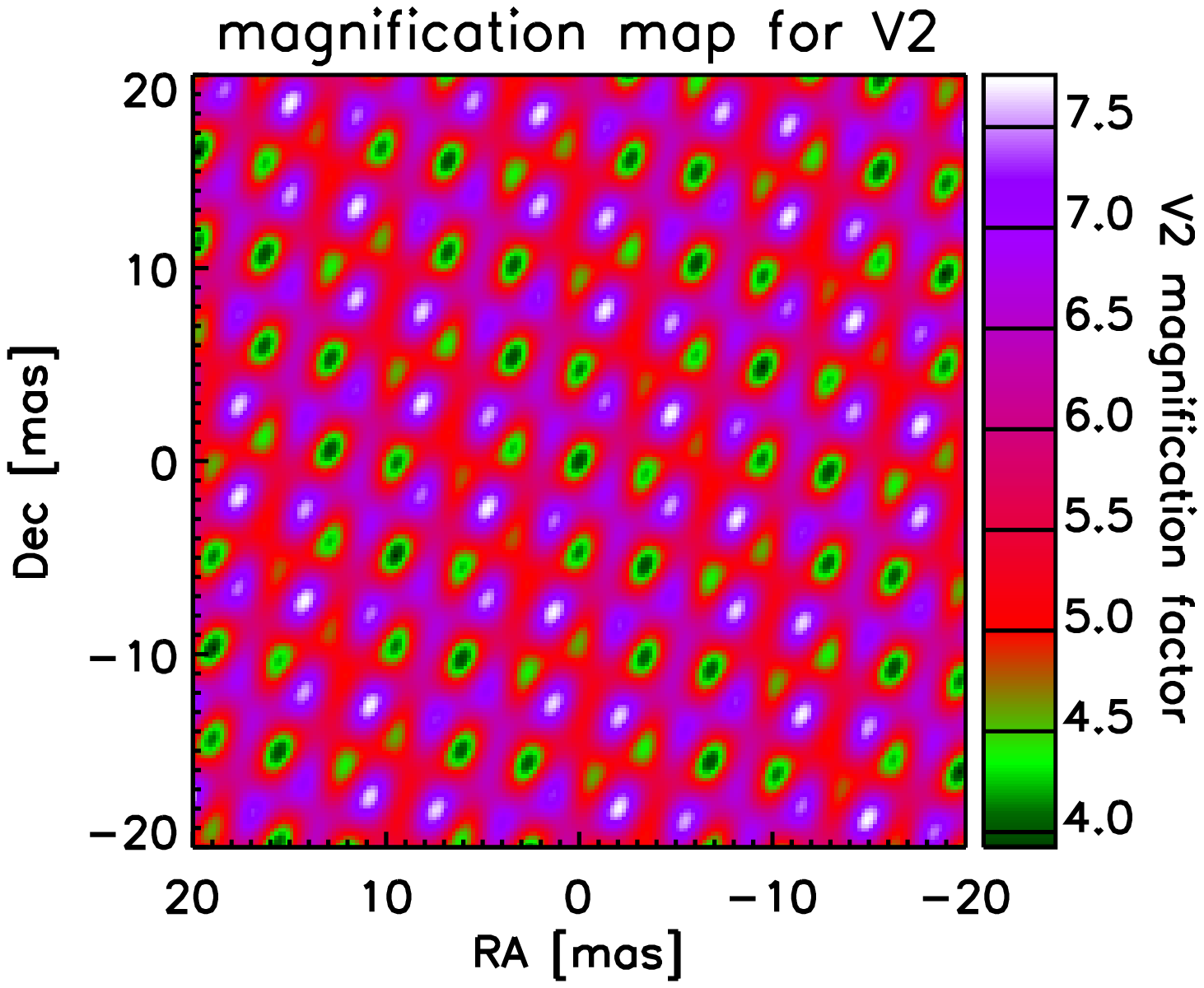}\quad
\includegraphics[scale=0.33]{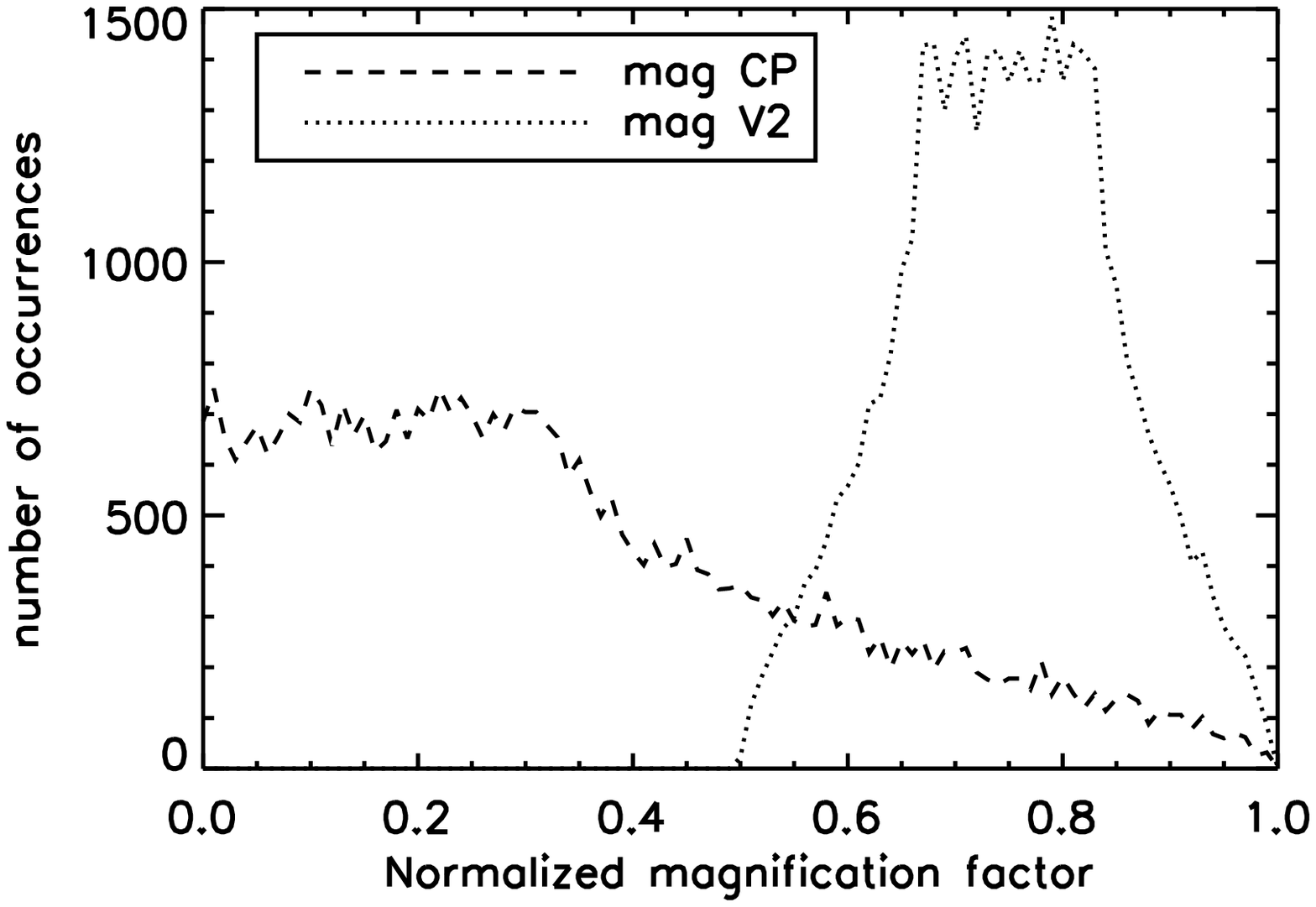}\\
 \includegraphics[scale=0.33]{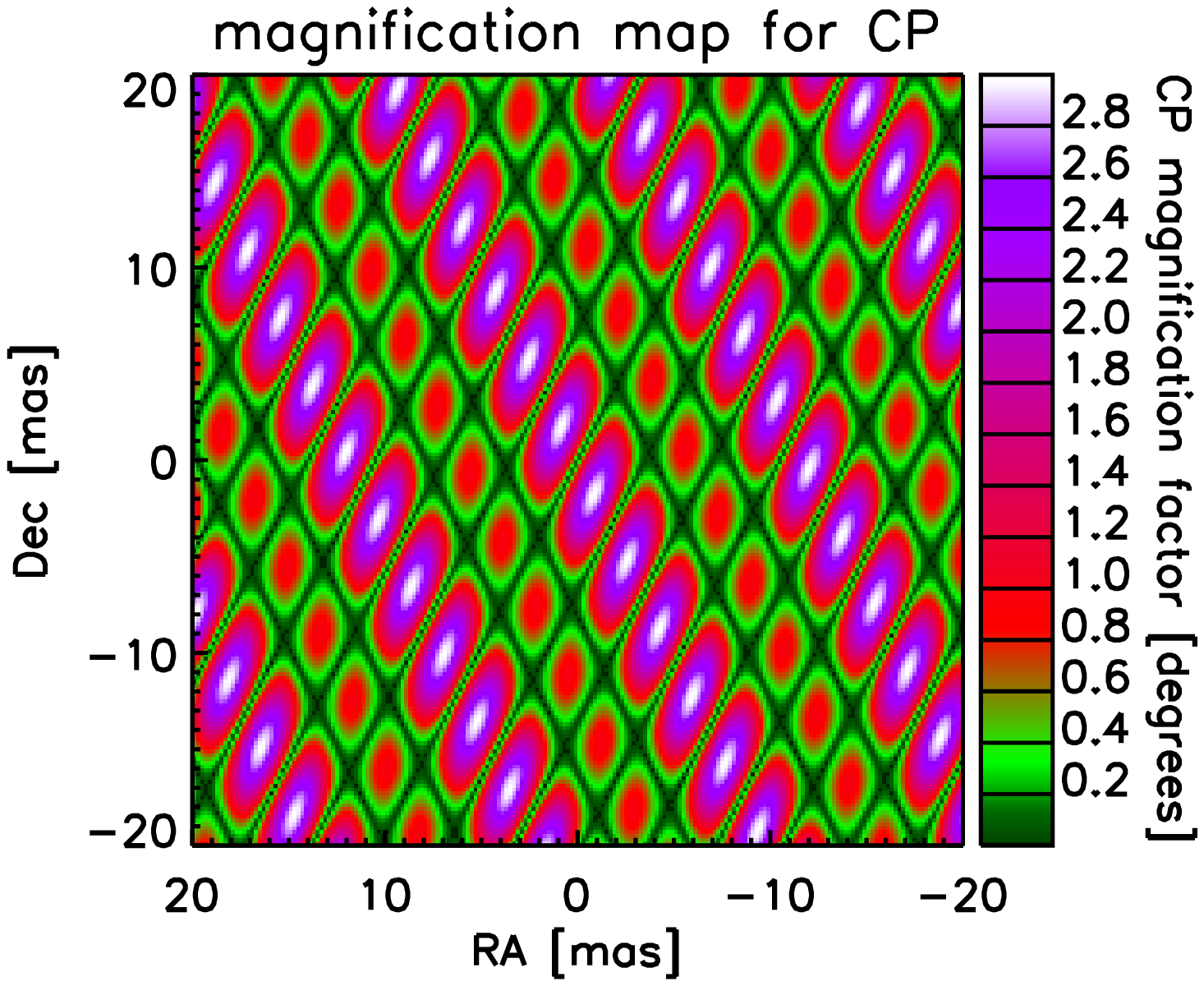}\quad
\includegraphics[scale=0.33]{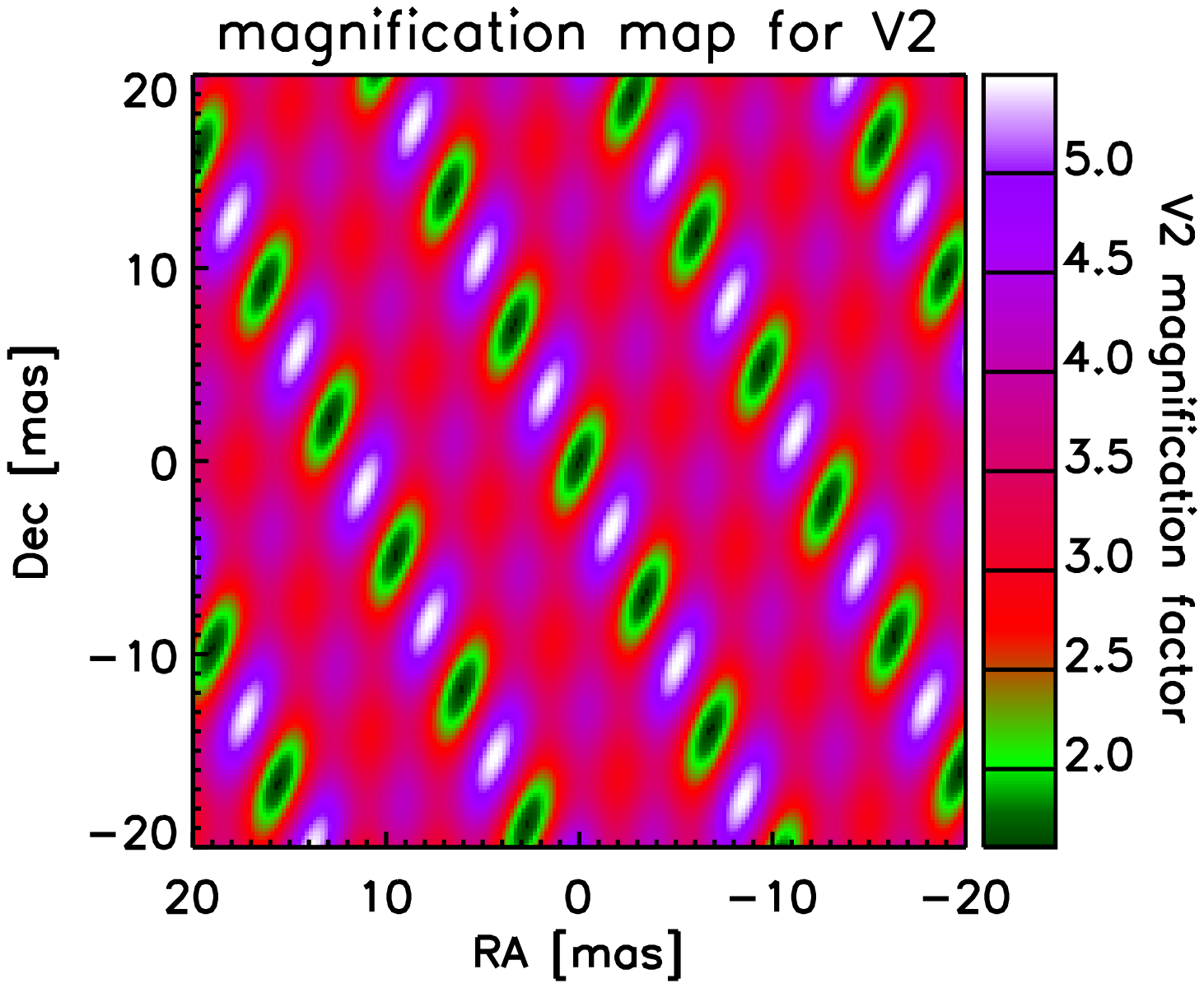}\quad
\includegraphics[scale=0.33]{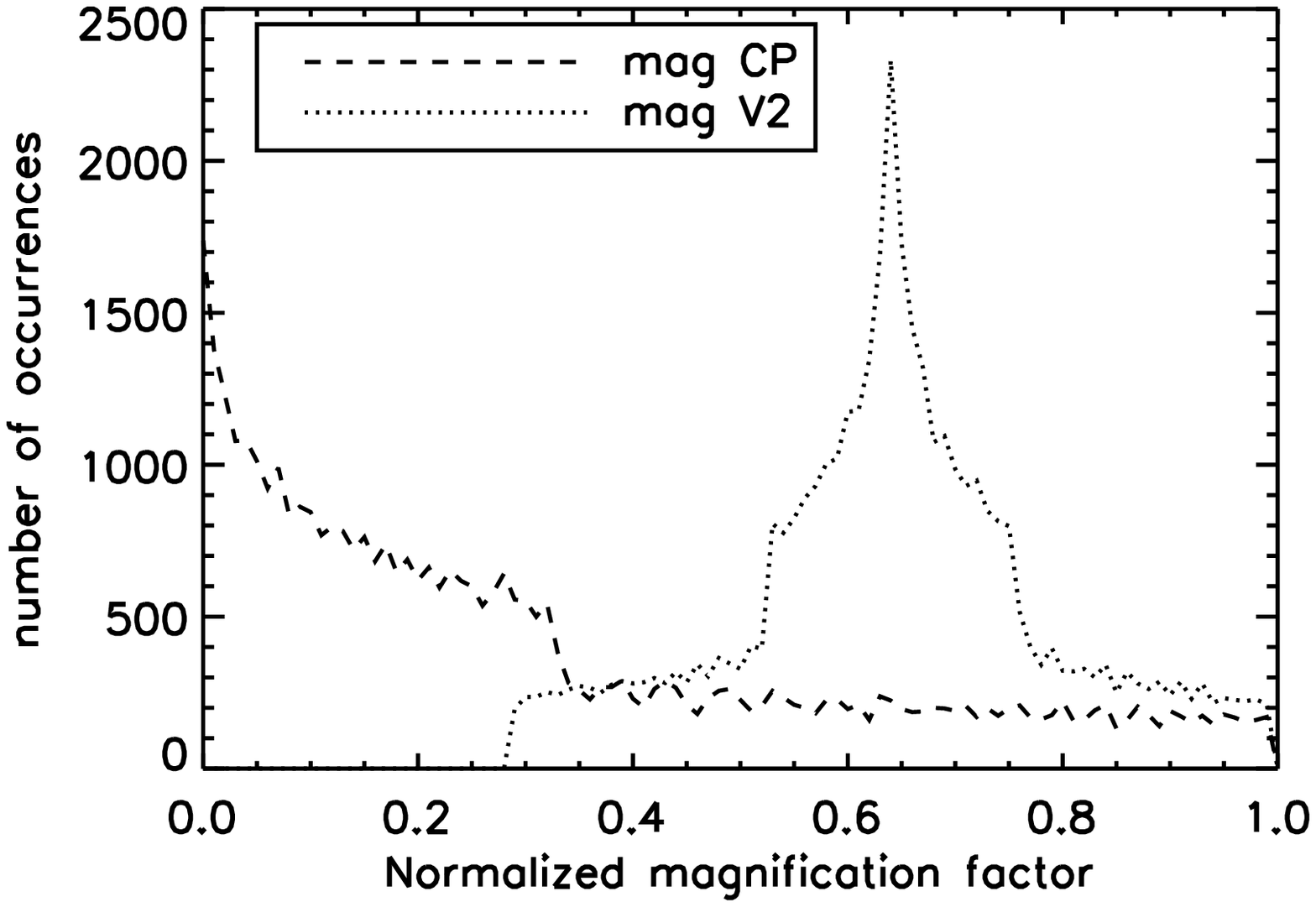}
\caption{Maps of the magnification factor in absolute value for the CP (left) and the $V^2$ (middle), together with the normalised histograms for both magnification maps, for two different telescope configurations: A1--K0--G1 (triangular configuration, top) and D0--G0--H0 (linear configuration, bottom).}
 \label{fig:magmaps}
\end{figure*}

Figure~\ref{fig:sensitivitydet} also gives interesting information on the sensitivities of the CP and $V^2$ individually. They show a similar median sensitivity level ($\sim 1$\%), but the situation changes when considering a completeness level of 90\%, for which the sensitivity of the CP (2.7\%) degrades much more than for the $V^2$ (1.6\%). This is somewhat surprising as the closure phases are generally thought to be more sensitive to the presence of faint companions. To investigate the origin of this unexpected behaviour, we make use of the magnification factor defined by \citet{LeBouquin12} for the CP,
\begin{equation}
m_{\rm CP} = sin(\alpha_{12})+sin(\alpha_{23})-sin(\alpha_{12}+\alpha_{23}) \; ,
\end{equation}
where $\alpha_{ij}=2\pi \vec{B}_{ij}\cdot\vec{\Delta}/ \lambda$, with $\vec{B}_{ij}$ the baseline vector, $\vec{\Delta}$ the apparent binary separation vector, and $\cdot$ the scalar product. This purely geometric factor captures the amplification of the CP created by an off-axis companion as a function of its position in the field of view. It is possible to define an equivalent magnification factor for the visibilities, assuming an unresolved primary star,
\begin{equation}
m_{V^2} = \sum_{ij}(1-V^2_{ij}) \; ,
\end{equation}
with $V^2_{ij}$ the squared visibility for baseline $ij$. This definition gives the cumulated drop of visibility created by the companion on all baselines; the larger the drop, the more conspicuous the companion. The magnification maps for two different three-telescope configurations of the VLTI sub-array are given in Fig.~\ref{fig:magmaps}. The absolute value of the magnification factor is plotted to reveal more clearly the blind spots (green-black regions). We can easily note that, in both configurations, the closure phase magnification map comprises many more regions where the magnification factor is close to zero. This is probably related to the fact that the magnification for the CP ranges between negative and positive values, while the magnification for the $V^2$ is always positive. The closure phase therefore shows many more blind spots, where the presence of a companion would not show up in the data. This becomes even clearer in the histograms plotted in Fig.~\ref{fig:magmaps} (right), where we see that most of the occurrences for $m_{\rm CP}$ are close to zero. Even though the CP are arguably more robust than the $V^2$ for the detection of a companion, the poor coverage of the field of view limits the completeness of the search at a given contrast level. We conclude that taking into account the $V^2$ is highly recommended when searching for faint companions, despite the possible presence of false positives in the combined $\chi^2$ related to circumstellar discs.

%__________________________________________________________________

\section{Conclusions} \label{sec:concl}

In this paper, we have systematically searched for companions around the 92 stars observed within the \textsc{Exozodi} survey, which aims to unveil the occurrence rate of bright exozodiacal discs around nearby main sequence stars using infrared interferometry. Based on our VLTI/PIONIER observations, five new companions are resolved around HD~4150, HD~16555, HD~29388, HD~202730, and HD~224392, although three of these stars (HD~4150, HD~16555 and HD~224392) were already suspected to be binaries based on Hipparcos astrometry, and one of the binaries (HD~16555) was independently resolved by speckle interferometry while we were carrying out our survey. All these companions happen to be detected around A-type stars. They have estimated spectral types ranging from K4V to A5V, assuming that they are all on the main sequence. The fact that such bright companions have remained undiscovered to date led us to discuss how our observations affect the current estimation of binary fraction around A-type stars. In particular, based on our discoveries, the fraction of visual binaries increases from about 34\% to about 47\%. These serendipitous discoveries suggest that a significant fraction of supposedly single A-type stars are still undetected binaries. The \textsc{Exozodi} data set also allowed us to study the sensitivity of PIONIER to off-axis companions in its survey mode (3 OBs per star), showing a typical contrast limit of 1\%. Finally, we conclude that using the squared visibilities together with the closure phases in the search for companions is crucial for maximising the completeness of the search.

\begin{acknowledgements}
The authors thank the French National Research Agency (ANR, contract ANR-2010 BLAN-0505-01, EXOZODI) for financial support. L.M.\ acknowledges the F.R.S.-FNRS for financial support through a FRIA PhD fellowship. We thank S.~Borgniet and A.-M.~Lagrange for helpful discussions on HD~224392 and on radial velocity surveys of early-type stars. This work made use of the Smithsonian/NASA Astrophysics Data System (ADS) and of the Centre de Donn\'ees astronomiques de Strasbourg (CDS).
\end{acknowledgements}

\bibliographystyle{aa} % style aa.bst
\bibliography{PIONIER_companions_rev} % your references Yourfile.bib

\end{document}